\documentclass[reprint,amsmath,amssymb,prl,]{revtex4-2}
\usepackage{braket}
\usepackage{graphicx}
\usepackage{dcolumn}
\usepackage{bm}
\usepackage{color} 
\usepackage{ulem}
\usepackage{tabularx}
\usepackage{xcolor}

\begin{document}
\title{Robust storage qubits in ultracold polar molecules
}
\author{Philip D. Gregory$^{1}$}
\author{Jacob A. Blackmore$^{1}$}
\author{Sarah L. Bromley$^{1}$}
\author{Jeremy M. Hutson$^{2}$}
\author{Simon L. Cornish$^{1}$}
 \email{s.l.cornish@durham.ac.uk}
\address{
\mbox{$^{1}$Joint Quantum Centre (JQC) Durham-Newcastle, Department of Physics,} \mbox{Durham University, Durham, United Kingdom, DH1 3LE.} \\
\mbox{$^{2}$Joint Quantum Centre (JQC) Durham-Newcastle, Department of Chemistry,} \mbox{Durham University, Durham, United Kingdom, DH1 3LE.} \\
}


\begin{abstract}
Quantum states with long-lived coherence are essential for quantum computation, simulation and metrology. The nuclear spin states of ultracold molecules prepared in the singlet rovibrational ground state are an excellent candidate for encoding and storing quantum information. However, it is important to understand all sources of decoherence for these qubits, and then eliminate them, in order to reach the longest possible coherence times. Here, we fully characterise the dominant mechanisms for decoherence of a storage qubit in an optically trapped ultracold gas of RbCs molecules using high-resolution Ramsey spectroscopy. Guided by a detailed understanding of the hyperfine structure of the molecule, we tune the magnetic field to where a pair of hyperfine states have the same magnetic moment. These states form a qubit, which is insensitive to variations in magnetic field. Our experiments reveal an unexpected differential tensor light shift between the states, caused by weak mixing of rotational states. We demonstrate how this light shift can be eliminated by setting the angle between the linearly polarised trap light and the applied magnetic field to a magic angle of $\arccos{(1/\sqrt{3})}\approx55^{\circ}$. This leads to a coherence time exceeding 6.9\,s (90\% confidence level).
Our results unlock the potential of ultracold molecules as a platform for quantum computation.
\end{abstract}

\maketitle

Quantum coherence is a key resource~\cite{streltsov2017}, underpinning many prominent applications in quantum science and technology. These range from precision tests of fundamental physics~\cite{Safronova2018}, quantum metrology~\cite{Giovannetti2011} and state-of-the-art atomic clocks~\cite{Ludlow2015} to quantum information processing~\cite{Ladd2010}, quantum simulation~\cite{Georgescu2014} and quantum thermodynamics~\cite{Vinjanampathy2016}. Understanding the limits on quantum coherence is therefore of fundamental interest and technological importance. Cooling matter into the ultracold regime leads to long interrogation times coupled with exquisite experimental control, enabling quantum coherence to be investigated  with incomparable precision.

Ultracold polar molecules~\cite{Carr2009,Bohn2017}  combine the rich internal structure associated with molecular vibration and rotation with access to controllable long-range dipole-dipole interactions. These properties have stimulated a diverse range of proposed applications spanning the fields of quantum computation~\cite{DeMille2002, Yelin2006, Ni2018, Sawant2020, Hughes2020}, quantum simulation~\cite{Barnett2006, Micheli2006, Gorshkov2011, Manmana2013}, quantum-state controlled chemistry~\cite{Krems2008,Balakrishnan2016,Hu2020}, and precision tests of fundamental physics~\cite{Zelevinsky2008, Hudson2011, ACME2018}. To realise many of these applications, we need to understand how to engineer long-lived quantum coherence in ultracold polar molecules.

In this work, we use high-precision Ramsey spectroscopy to investigate the sources of decoherence in an optically trapped ultracold gas of $^{87}$Rb$^{133}$Cs molecules (hereafter RbCs). 
We focus on superpositions of nuclear spin states of the singlet rovibrational ground state. Such superpositions are expected to be relatively insensitive to magnetic dephasing, as the magnetic moments of the nuclear spins are small in comparison to electronic magnetic moments. Furthermore, the nuclear spin states are expected to experience near-identical AC Stark shifts in an optical trap, so that dephasing associated with the nonuniform optical potential is also suppressed. These properties point to the possibility of long-lived coherence and make the nuclear spin states of ultracold polar molecules excellent candidates for robust storage qubits in quantum computing architectures~\cite{Yelin2006, Ni2018,Park2017}. In such proposals, gate operations may be performed using dipolar-exchange interactions~\cite{Barnett2006, Gorshkov2011} following microwave excitation to an excited rotational state, while single-qubit rotations can be performed using two-photon microwave pulses~\cite{Neyenhuis2012, Will2016, Gregory2016,Blackmore2020pccp}. 
Here, we demonstrate coherence times exceeding 6.9 s (90\% confidence level) for the storage qubit, paving the way for the use of ultracold molecules as a platform for quantum computation.

\begin{figure*}[t]
    \centering
    \includegraphics[width=\textwidth]{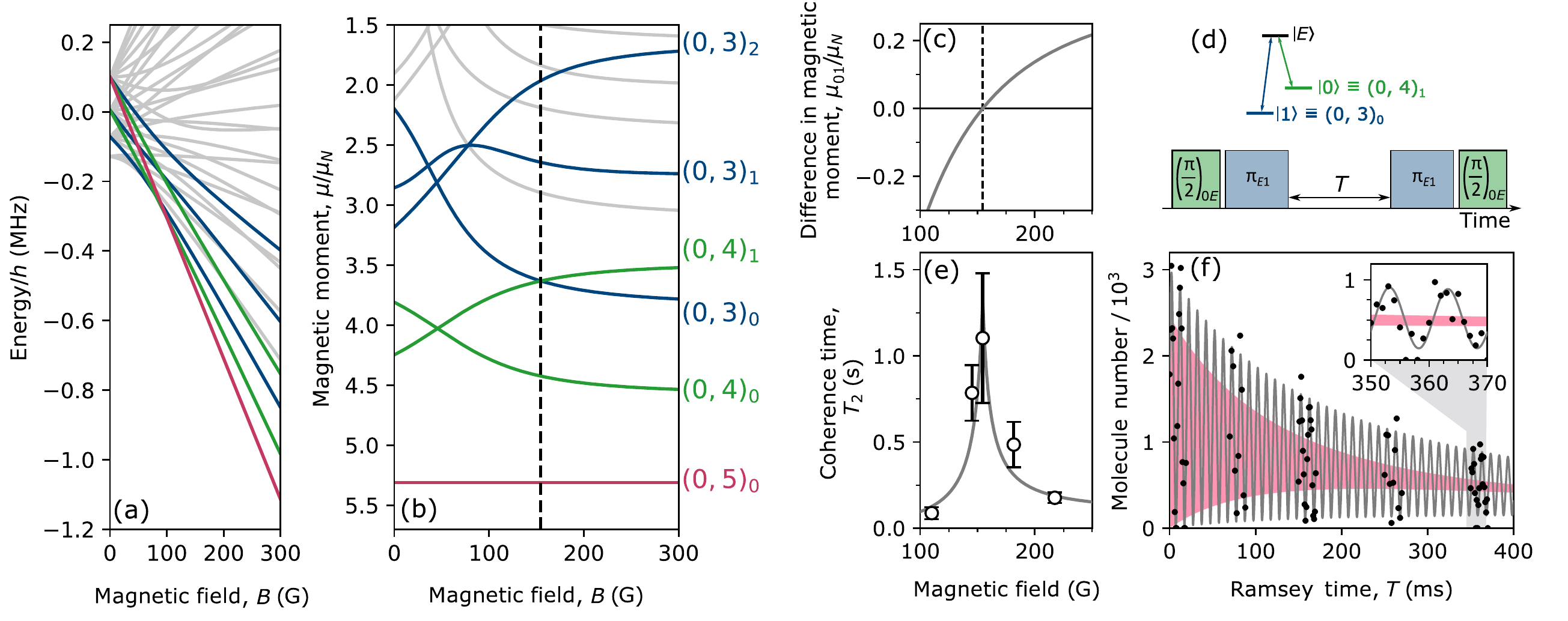}
    \caption{{\bf Effect of magnetic fields on the coherence of the storage qubit.} {\bf (a)}~Zeeman structure of the nuclear spin states in the rotational ground state of RbCs. {\bf (b)}~Magnetic moments $\mu/\mu_\mathrm{N}$ of the states as a function of magnetic field $B$. We construct the qubit from the states \mbox{$\ket{0}\equiv(0,4)_1$} and $\ket{1}\equiv(0,3)_0$, which are chosen as they possess identical magnetic moments when $B=154.524$\,G (indicated by the vertical dashed line). {\bf (c)}~The differential magnetic moment $\mu_{01}=\mu_{\ket{1}} - \mu_{\ket{0}}$ as a function of magnetic field. {\bf (d)}~Configuration of states and microwave pulse sequence used to perform Ramsey spectroscopy. The energy separation between the qubit states and the rotationally excited state $\ket{E}$ is $2B_v\approx980$\,MHz. {\bf (e)}~Measured coherence time $T_2$ as a function of magnetic field. The line shows a model for the decoherence as described in the text. Fitting to the results indicates a magnetic field variation of 34(5)\,mG over the course of the measurement contributes to the observed decoherence, and a peak coherence time $T^*_2=1.3(4)$\,s when $\mu_{01}\approx0$. {\bf (f)}~Example Ramsey measurement performed at $B=154.50$\,G, where $\mu_{01}\approx0$. The $y$-axis indicates the number of molecules remaining in state $\ket{0}$ measured as a function of the Ramsey time $T$. The red shaded region indicates the maximum and minimum of the Ramsey fringes observed when $B=217.39$\,G, where $\mu_{01}\approx0.17\,\mu_\mathrm{N}$ and the coherence time is correspondingly much shorter. The trap light has polarisation $\beta=0^{\circ}$ and intensity $I=15.8$\,kW\,cm$^{-2}$ for all measurements shown.}
    \label{fig:Theory}
\end{figure*}

To begin, we seek to identify pairs of nuclear spin states with identical magnetic moments that connect to a common excited rotational state, by calculating the rotational and hyperfine structure of the RbCs molecule in externally applied magnetic and  optical fields~\cite{Gregory2016, Gregory2017, Blackmore2020, Blackmore2020pccp}. 
We construct the Hamiltonian (see Methods) in a fully uncoupled basis set $\ket{N, M_N} \ket{i_\mathrm{Rb}, m_\mathrm{Rb}} \ket{i_\mathrm{Cs}, m_\mathrm{Cs}}$, where $N$ represents the angular momentum of the molecule with its projection along the quantisation axis $M_N$, and $i_\mathrm{Rb}=3/2, i_\mathrm{Cs}=7/2$ denote the nuclear spins of Rb and Cs respectively, with their projections $m_\mathrm{Rb}, m_\mathrm{Cs}$. 
However, typical magnetic fields in our experiments are not high enough to decouple the rotational and nuclear angular momenta. Even when the laser is polarised along the magnetic field, the only good quantum number that can be used to describe a given hyperfine sublevel is $M_F=M_N+m_\mathrm{Rb}+m_\mathrm{Cs}$.
As this is not sufficient to identify a given hyperfine state uniquely, we label the states by $(N, M_F)_k$ where $k$ is an index counting up the states in order of increasing energy, such that $k=0$ is the lowest-energy state for given values of $N$ and $M_F$. There are 32 nuclear spin states in the $N=0$ rotational ground state of RbCs, with energies $E$ shown in Fig.~\ref{fig:Theory}(a). The magnetic moments $\mu=dE/dB$ for a selection of the states are plotted as a function of magnetic field $B$ in Fig.~\ref{fig:Theory}(b). There are a number of state combinations which display crossings where the difference in magnetic moments is zero.
These crossings indicate turning points in the energy difference between states, where the energy difference becomes insensitive to magnetic field noise.
Our experiment produces an optically trapped ultracold gas of RbCs molecules by association from a pre-cooled atomic mixture~\cite{Molony2014, Molony2016}, using a procedure (see Methods) that initialises the molecules in the state $(0,4)_1$; full state compositions are given in the Supplementary Information. For simplicity, we therefore select the qubit states to be $\ket{0}\equiv(0,4)_1$ and $\ket{1}\equiv(0,3)_0$ which are predicted to have the same magnetic moment when $B=154.524$\,G, as shown in~Fig.~\ref{fig:Theory}(c), where we plot $\mu_{01}=\mu_{\ket{1}} - \mu_{\ket{0}}$.

\begin{figure*}[t]
    \centering
    \includegraphics[width=\textwidth]{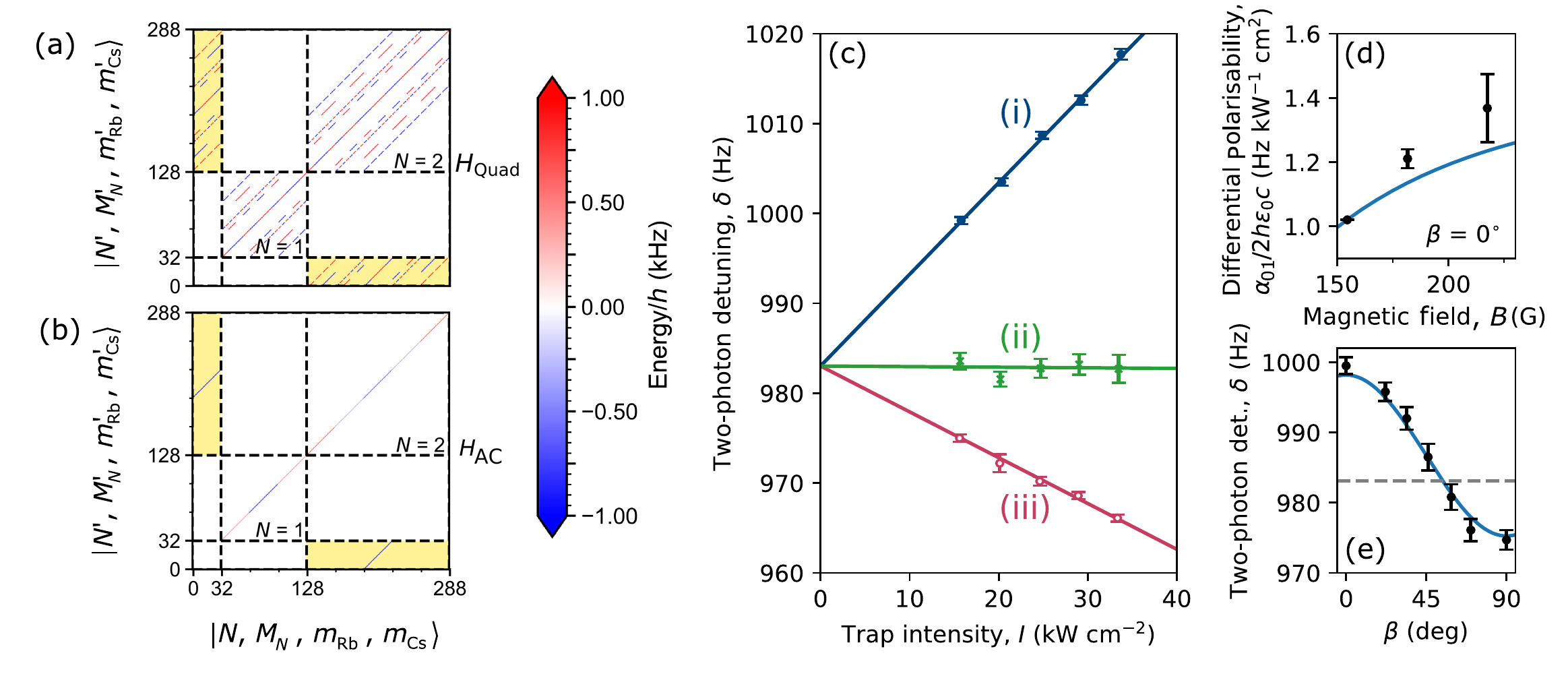}
    \caption{{\bf Differential tensor light shifts between nuclear spin states in the rotational ground state.} We measure the differential light shifts between \mbox{$\ket{0}\equiv(0,4)_1$} and $\ket{1}\equiv(0,3)_0$ by Ramsey spectroscopy. We show the Hamiltonian matrix elements for {\bf (a)}~the nuclear electric quadrupole interaction $H_\mathrm{quad}$ and {\bf (b)}~the AC Stark effect $H_\mathrm{AC}$ graphically for the uncoupled $\ket{N,M_N,m_\mathrm{Rb},m_\mathrm{Cs}}$ basis. The states are split by dashed lines into groups according to $N$ and by minor ticks into multiples of 32 basis states with the same $M_N$. The block diagonal elements in $N$ are labelled for $N>0$. The color coding indicates the value of the matrix element in units of frequency. Note the off-diagonal elements connecting states with $\Delta N=2$ shaded in yellow, which lead to tensor light shifts proportional to the anisotropic polarisability. The mathematical expressions describing the matrix elements are given in the Supplementary Information. $H_\mathrm{AC}^{(2)}$ is calculated for an intensity of $16~\mathrm{kW\,cm^{-2}}$ and a polarisation angle of $\beta=0^\circ$. {\bf (c)}~Two-photon detuning $\delta$ as a function of the trap intensity $I$. The linear polarisation of each beam is set to an angle $\beta$ of (i)\,0$^{\circ}$ (ii)\,55$^{\circ}$ (iii)\,90$^{\circ}$ with respect to a 154.50\,G magnetic field. The coloured lines indicate a fit to the results, following the model given in Eq.~\ref{eq:ACStarkFit}. We find $\alpha^{(2)}/(4\pi\epsilon_0)=545(4)\,a_0^{3}$, and $\delta_0=983.0(2)$\,Hz. {\bf (d)}~Differential polarisability between the states as a function of magnetic field, measured for $\beta=0^{\circ}$. The line is the expectation from the molecular Hamiltonian (see Methods). {\bf (e)}~$\delta$ measured in a single beam of the dipole trap with fixed intensity $I=15.3$\,kW\,cm$^{-2}$, as a function of $\beta$. The dashed horizontal line indicates $\delta_0$. The solid line is calculated using Eq.\ref{eq:ACStarkFit} with the parameters found in~(c).}
    \label{fig:ACStark}
\end{figure*}

We investigate the dependence of the coherence time $T_2$ on $\mu_{01}$ by performing a series of Ramsey measurements to measure $T_2$ as a function of $B$ (see Fig.~\ref{fig:Theory}(d) and Methods). The results are shown in Fig.~\ref{fig:Theory}(e). We observe the longest coherence time when the difference in magnetic moments between the two states is zero, as expected. We fit the magnetic field variation of $T_2$ with
\begin{equation}
T_2 = \left(\frac{|\mu_{01}|~\Delta B}{h} + \frac{1}{T^*_2}\right)^{-1},
\label{eq:MagneticFieldFit}
\end{equation}
where $\Delta B$ and $T^*_2$ are fitting parameters (see the Supplementary Information for a derivation of Eq.~\ref{eq:MagneticFieldFit}). $\mu_{01}$ is calculated from the molecular Hamiltonian as shown in Fig.~\ref{fig:Theory}(c). $\Delta B$ describes the magnitude of variation in magnetic field over the duration of the measurement which contributes to the decoherence. 
We find $\Delta B = 34(5)$\,mG, which is consistent with the expected stability of the magnetic fields in our experiments. The term $T^*_2$ accounts for other sources of decoherence, which we show below to be dominated by differential tensor light shifts. For these measurements we find $T^*_2=1.3(4)$\,s, for trap light polarised with $\beta=0^{\circ}$ and intensity $I=15.8$\,kW\,cm$^{-2}$. Fig.~\ref{fig:Theory}(f) shows Ramsey fringes recorded close to the $\mu_{01}=0$ condition and contrasts the behaviour with that seen at $B=217.39$\,G where, although the difference in magnetic moments is still small, magnetic dephasing limits the coherence time. We estimate that the limit on the coherence time at $B=154.52$\,G due to magnetic field noise of $\Delta B=35$\,mG is $\sim2.0\times10^3$\,s (see Supplementary Information).

To show that the remaining decoherence $T_2^*$ is dominated by differential tensor light shifts, we perform a series of Ramsey measurements using different optical trap intensities. Each Ramsey measurement allows us to precisely determine the difference in energy between $\ket{0}$ and $\ket{1}$. For these experiments the two microwaves fields differ in frequency by 76\,kHz, and so by measuring the frequency of the Ramsey fringes $\delta$, we determine the difference in energy between the states equal to $h\times(76\,\mathrm{kHz}+\delta)$. The sign of $\delta$ is found by comparison with additional Ramsey measurements with intentionally different two-photon detunings. We measure $\delta$ for a range of trap laser intensities, and find an intensity-dependent energy shift between the two states as shown in Fig.~\ref{fig:ACStark}.

The differential light shift arises from terms off-diagonal in $N$ which cause mixing between states with the same parity. 
The largest contributions to the light shift are second-order terms
\begin{equation}
\bra{N=0, M_N=0}H_\mathrm{AC}\ket{2, 0}\bra{2, 0}H_\mathrm{quad}\ket{0, 0},
\label{eq:MatElements}
\end{equation}
where $H_\mathrm{AC}$ and $H_\mathrm{quad}$ represent the the AC Stark and nuclear electric quadrupole interactions, respectively (see Methods). 
The matrix elements of $H_\mathrm{AC}$ and $H_\mathrm{quad}$ are shown graphically in Figs.~\ref{fig:ACStark}(a) and (b).
These second-order terms lead to components with $N>0$ in the state composition of $\ket{0}$ and $\ket{1}$ with coefficients $<10^{-5}$. 
This results in tensor light shifts in the rotational ground state which scale with the anisotropic polarisability $\alpha^{(2)}$~\cite{Gregory2017} and depend on $M_F$ and the laser polarisation. This is analogous to the tensor light shifts that arise in ground-state alkali atoms due to hyperfine structure~\cite{Sandars1967}. The terms in Eq.~\ref{eq:MatElements} are all diagonal in $M_N$ and are proportional to $P_2(\cos\beta)=\frac{1}{2}(3\cos^{2}\beta - 1)$, where $\beta$ is the angle between the linearly polarised electric field of the trap light and the applied magnetic field which forms the quantisation axis. As a result, the light shift changes the observed two-photon detuning according to
\begin{equation}
\delta=(\alpha_{01} I)/2 h \epsilon_0 c + \delta_0,
\label{eq:ACStarkFit}
\end{equation}
where 
\begin{equation}
\alpha_{01} = X(B) \alpha^{(2)} P_2(\cos\beta),
\label{eq:ACStark}
\end{equation} 
is the difference in the effective differential polarisability between the states. Here, $I$ is the average intensity experienced by the molecules, $\delta_0$ is the two-photon detuning in free space, and $X(B)$ is a numerical factor which is determined from the molecular Hamiltonian and depends upon the magnetic field.

In addition to the tensor light shift, each state also experiences a much larger scalar light shift. However, as the scalar shift is identical for all states, this does not contribute to decoherence. The largest differential tensor light shift we measure is $1.01995(6)$\,Hz\,kW$^{-1}$\,cm$^{2}$, for $\beta=0^{\circ}$. This is caused by individual tensor lights shifts for each of the states which we calculate to be \mbox{$-1.45$\,Hz\,kW$^{-1}$\,cm$^{2}$} for~$\ket{0}$, and \mbox{$-2.47$\,Hz\,kW$^{-1}$\,cm$^{2}$} for~$\ket{1}$. In contrast the scalar light shift for both states is \mbox{$-41.2$\,kHz\,kW$^{-1}$\,cm$^{2}$}; this is 4 orders of magnitude larger than the tensor light shifts.

We compare our calculations with the behaviour observed in experiments. For a magnetic field of $B\approx154.50$\,G where magnetic decoherence is minimised, 
we calculate the prefactor $X(B)=4.00(4)\times10^{-5}$. We plot the differential light shift measured at this magnetic field in the optical trap for fixed laser polarisations $\beta=0^{\circ}, 54^{\circ}$, 90$^{\circ}$ in Fig.~\ref{fig:ACStark}(c). The solid lines indicate a fit to the results using Eq.~\ref{eq:ACStarkFit}, with $\alpha^{(2)}$ and $\delta_0$ as free parameters. We find excellent agreement between our model and the experiment, with \mbox{$\alpha^{(2)}/4\pi\epsilon_0=545(4)\,a_0^3$} and \mbox{$\delta_0=983.0(2)$\,Hz}. The uncertainties shown are the statistical uncertainties found in the fitting. Additional systematic uncertainties in $\alpha^{(2)}$ are given in the Supplementary Information.  
The value of $\delta_0$ indicates the two-photon detuning in free space, and so we determine the free-space energy difference between the states of $h\times(76\,\mathrm{kHz} + \delta_0) = h\times76.983\,0(2)$\,kHz. This is in excellent agreement with a calculation from the molecular Hamiltonian which predicts an energy difference between the states of $h\times77.0(7)$\,kHz, where the uncertainty results from the current precision with which the strength of the scalar nuclear spin-spin interaction ($c_4$) and the magnitude of the nuclear magnetic moments are known for RbCs~\cite{Gregory2016, Blackmore2020pccp}.

To test our understanding of the origin of the differential light shift further, we explore different magnetic fields as shown in Fig.~\ref{fig:ACStark}(d). For higher magnetic fields, the measurements are performed for $\beta=0^{\circ}$ only. The increased uncertainties arise from the magnetic dephasing restricting the measurement time. The variation with magnetic field arises from the numerical prefactor $X(B)$ in Eq.~\ref{eq:ACStark}.  We find good qualitative agreement between theory and experiment, with $\alpha_{01}$ rising with magnetic field. For calculations over a broader range of magnetic fields see the Supplementary Information. Our theory appears to underestimate the increase in $\alpha_{01}$ slightly. We attribute the discrepancy to uncertainties in the parameters of the molecular Hamiltonian which combine in a non-trivial way in the calculation of $X(B)$. Ramsey measurements of the type presented here should permit further refinement of these parameters. This will be the focus of future work.

The tensor light shifts we observe are proportional to $P_2(\cos\beta)$. This allows us to engineer a magic polarisation trap, as $P_2(\cos\beta)=0$ for the magic angle~$\beta_\mathrm{magic}=\arccos{\sqrt{1/3}}\approx55^{\circ}$. We experimentally verify this angle dependence in Fig.~\ref{fig:ACStark}(e) using a single beam of the dipole trap. We see that the polarisation dependence of the experimentally measured $\delta$ is well described by our model and that $\delta\approx\delta_0$ when $\beta\approx55^{\circ}$, indicating that the tensor light shift is zero. This is further confirmed using the measurements in Fig.~\ref{fig:ACStark}(c), where all $\delta$ measured in the trap for $\beta=55^{\circ}$ are consistent with the free-space value, and the gradient of $\delta$ as a function of $I$ is zero.

\begin{figure*}[t]
    \centering
    \includegraphics[width=\textwidth]{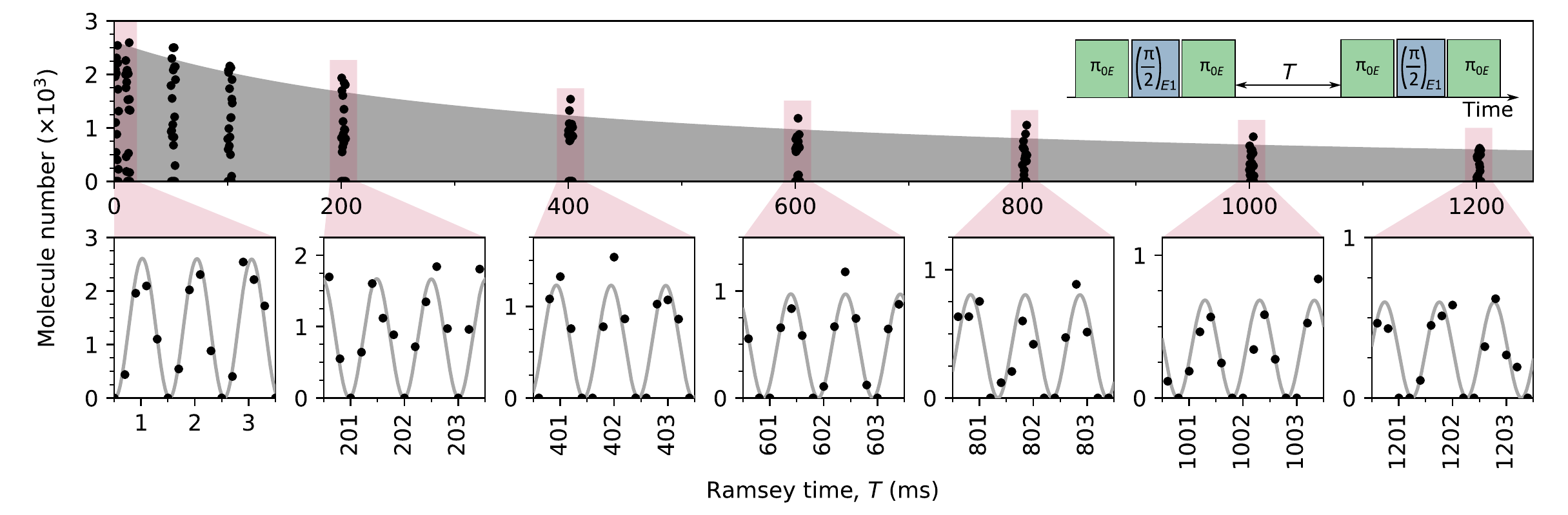}
    \caption{{\bf Robust coherence of the storage qubit.} Ramsey experiment with the qubit states $\ket{0}\equiv(0,4)_1$ and $\ket{1}\equiv(0,3)_0$, using the optimal configuration $B\approx154.50$\,G, $\beta\approx55^{\circ}$. The $y$-axis indicates the number of molecules remaining in $\ket{0}$, following the Ramsey sequence shown inset, as a function of the hold time $T$. The shaded region in the  upper plot indicates the maximum and minimum of the Ramsey fringes as a function of time; the spacing of the fringes is too small to plot at this scale. The lower plots show the Ramsey fringes observed at 400\,ms intervals.    
    }
    \label{fig:BestRamsey}
\end{figure*}

The optimal configuration to maximise coherence time is where $\mu_{01}=0$ and $\alpha_{01}=0$, which is realised in our experiments for $B\approx154.5$\,G and $\beta\approx55^{\circ}$. We perform a Ramsey experiment using these optimal parameters as shown in Fig.~\ref{fig:BestRamsey}. The maximum Ramsey time available is limited by collisional loss of the molecules~\cite{Gregory2019,Gregory2020} with $T_1=0.61(4)$ seconds, which reduces our signal at long times. We measure Ramsey fringes out to $T=1.2$\,s with no evidence of decoherence. These results are consistent with $T^*_2>6.9$\,s (90\,\% confidence level, see Supplementary Information), an order of magnitude longer than any previous work~\cite{Park2017}. Using this measurement we find the energy difference between the states in the trap to be $h\times76,982.733(16)$\,Hz; this is a precision of 1 part in $10^{7}$. 

Our measurements do not indicate any other detectable mechanisms for decoherence. We see no evidence for collisional energy shifts, which would be observed by a change in the energies of the states when the density reduces over the course of each Ramsey measurement (see Supplementary Information). This is consistent with previous observations~\cite{Park2017}, and the absence of collisional energy shifts or decoherence may be expected as short-range collisions in the gas lead to loss of molecules with high probability~\cite{Ye2018, Gregory2019, Gregory2020, Croft2020}. Measurements of the coherence out to longer times will require confinement of the molecules to a 3D optical lattice~\cite{Zhu2014} , optical tweezers~\cite{Anderegg2019, Cheuk2020, Zhang2020}, or the use of alternative trapping techniques such as a blue-detuned optical trap~\cite{Bause2020} to avoid losses from the optical excitation of two-molecule collision complexes~\cite{Gregory2020, Liu2020}. The creation of controlled arrays of molecules is also a key component of the proposed quantum computing protocols where storage qubits have applications; our method of using a magic-polarisation trap is compatible with the confinement of molecules to arrays of optical tweezers or a 3D optical lattice~\cite{Kotochigova2010}.

In conclusion, we have demonstrated a robust storage qubit constructed from the nuclear spin states of ultracold RbCs molecules. We have shown how magnetic dephasing can be eliminated by tuning the magnetic field to where the states have identical magnetic moments. This revealed an unexpected differential tensor light shift due to weak mixing of rotational states of the same parity, which caused decoherence due to the nonuniform optical potential. We have shown how to eliminate these light shifts by setting the linear polarisation of the trap light to a magic angle $\beta_\mathrm{magic}=\arccos{\sqrt{1/3}}\approx55^{\circ}$ with respect to the magnetic field. Our optimal configuration leads to Ramsey fringes which persist for~$T^*_2>6.9$\,s (90\,\% confidence level), at least an order of magnitude improvement over the previous state of the art. Our findings are broadly applicable to all $^{1}\Sigma$ molecules in their rovibrational ground state, including the range of bialkali molecules currently under investigation. 
Our work demonstrates the implementation of robust storage qubits, which will be essential in future high-fidelity quantum computing architectures using controllable arrays of ultracold polar molecules.

\section*{Methods}

\subsection*{Hamiltonian used to calculate the rotational and hyperfine structure}

We calculate the energy level structure of RbCs in the electronic and vibrational ground state by diagonalising the relevant Hamiltonian. We extract the energy levels and eigenstates of the Hamiltonian by numerical diagonalisation.
The hyperfine constants for all of our calculations are given in the Supplementary Information. In the presence of externally applied magnetic and off-resonant optical fields, the Hamiltonian ($H_{\mathrm{RbCs}}$) can be decomposed into rotational ($H_\mathrm{rot}$), hyperfine ($H_\mathrm{hf}$), Zeeman ($H_\mathrm{Zeeman}$), and AC Stark ($H_\mathrm{AC}$) components~\cite{Brown&Carrington,Aldegunde2008}:
\begin{equation}
    H_\mathrm{RbCs} = H_\mathrm{rot}+H_\mathrm{hf}+H_\mathrm{Zeeman}+H_\mathrm{AC}.
    \label{eq:Hamiltonian}
\end{equation}
The rotational contribution 
\begin{equation}
\label{eqn:Rotational}
    H_\mathrm{rot} = B_v \bm{N}^2 -D_v \bm{N}^2\cdot\bm{N}^2,
\end{equation}
 is defined by the rotational angular momentum operator~$\bm{N}$, and the rotational and centrifugal distortion constants, $B_{v}$ and $D_{v}$. The hyperfine contribution consists of four terms
 \begin{equation}
    	H_\mathrm{hf}= H_\mathrm{quad} + H^{(0)}_{II} + H^{(2)}_{II} + H_{NI},
\end{equation}
where
\begin{subequations}
	\begin{gather}
	H_\mathrm{quad} = \sum_{j=\mathrm{Rb}, \mathrm{Cs}} e\boldsymbol{Q}_{j} \cdot \boldsymbol{q}_{j},\label{seq:NuclearQuad}\\
	H^{(0)}_{II} =c_{4} \boldsymbol{I}_{\mathrm{Rb}} \cdot \boldsymbol{I}_{\mathrm{Cs}},\\
	H^{(2)}_{II}=-c_{3}\sqrt{6}\boldsymbol{T}^2(C)\cdot \boldsymbol{T}^2\left(\boldsymbol{I}_\mathrm{Cs},\boldsymbol{I}_\mathrm{Rb}\right)\\
	H_{NI} = \sum_{j=\mathrm{Rb}, \mathrm{Cs}} c_{j} \boldsymbol{N} \cdot \boldsymbol{I}_{j}.
	\end{gather}
\end{subequations}
$H_\mathrm{quad}$ represents the interaction between the nuclear electric quadrupole of nucleus $j$ ($e\boldsymbol{Q}_j$) with the electric field gradient at the nucleus ($\boldsymbol{q}_j$). $H^{(0)}_{II}$ and $H^{(2)}_{II}$ are the scalar and tensor nuclear spin-spin interactions, with strengths governed by the coefficients $c_4$ and $c_3$. The second-rank tensors $\boldsymbol{T}^2$ describe the angular dependence and anisotropy of the interactions~\cite{Aldegunde2017}. $H_{NI}$ is the interaction between the nuclear magnetic moments and the magnetic field generated by the rotating molecule and has a coupling constant $c_j$ for each of the two nuclei.   

The Zeeman contribution to the Hamiltonian describes interaction of the rotational and nuclear magnetic moments with the external magnetic field ($\boldsymbol{B}$) and is
\begin{equation}
    \begin{split}
    H_\mathrm{Zeeman} =
    &-g_{\mathrm{r}} \mu_{\mathrm{N}} \boldsymbol{N} \cdot \boldsymbol{B} \\
    &-\sum_{j=\mathrm{Rb}, \mathrm{Cs}} g_{j}\left(1-\sigma_{j}\right) \mu_{\mathrm{N}} \boldsymbol{I}_{j} \cdot \boldsymbol{B}.\label{seq:Zeeman}
    \end{split}
\end{equation}
The first term accounts for the magnetic moment generated by the rotation of the molecule, characterised by the rotational $g$-factor $g_\mathrm{r}$. The second term accounts for the nuclear spin contributions, characterised by the nuclear $g$-factors $g_j$ shielded isotropically by the factor $\sigma_j$~\cite{Aldegunde2008}. In both terms $\mu_\textrm{N}$ is the nuclear magneton. For our analysis we designate the axis of the magnetic field $\boldsymbol{B}$ as the space-fixed $z$ axis and its magnitude as $B$. 

The AC Stark effect arises from the interaction of an off-resonant oscillating electric field $\boldsymbol{E}_\mathrm{AC}$ with the frequency-dependent molecular polarisability tensor $\boldsymbol{\alpha}$~\cite{Gregory2017}, and has a contribution to the Hamiltonian
\begin{equation}
    H_\mathrm{AC}= -\frac{1}{2} \boldsymbol{E}_\mathrm{AC}\cdot\boldsymbol{\alpha}\cdot\boldsymbol{E}_\mathrm{AC}.\label{seq:AC_Stark}
\end{equation}

The terms $H_\mathrm{quad}$, $H_{II}^{(2)}$, and $H_\mathrm{AC}$ all have components which are off-diagonal in $N$, and therefore contribute to the tensor light shifts which we observe. To be explicit, the matrix elements for these terms are included in Supplementary Information.

\subsection*{Production of ultracold RbCs molecules}

We produce ground-state RbCs molecules from an optically trapped ultracold mixture of \textsuperscript{87}Rb and \textsuperscript{133}Cs atoms using a two-step process. First, we use magnetoassociation on an interspecies Feshbach resonance at 197\,G~\cite{Koeppinger2014}. Following this, the remaining atoms are removed from the trap using the Stern-Gerlach effect. Second, the magnetic field is set to 181.6\,G, where the molecules are transferred to a single hyperfine sub-level of the $X^1\Sigma\,(v=0, N=0)$ rovibrational ground state using stimulated Raman adiabatic passage (STIRAP)~\cite{Molony2014,Gregory2015,Molony2016}. We set the STIRAP to initialise the molecules in~$\ket{0}\equiv(0,4)_1$, and the transfer is performed in free space to avoid spatially varying AC Stark shifts which otherwise limit the efficiency~\cite{Molony2016}. Following STIRAP, the molecules are recaptured in a crossed optical dipole trap at $\lambda=1550~\mathrm{nm}$; see the Supplementary Information for details. Both beams are linearly polarised at an angle $\beta$ with respect to the applied magnetic field.  The molecules have a typical temperature of 0.7\,$\mu$K, and a peak density of $\sim 1\times10^{11}\,$cm$^{-3}$. We detect molecules by reversing the creation process and imaging the resulting atomic clouds. As such, we only image molecules which occupy~$\ket{0}$.

\subsection*{Ramsey measurement protocol}

To measure the coherence time of the qubit, we perform Ramsey spectroscopy. To couple the qubit states, we use two microwave fields to form a 3-level lambda system, where both qubit states are coupled to a common rotationally excited state $\ket{E}$, which is chosen to have significant transition dipole moment to both $\ket{0}$ and $\ket{1}$. The excited states used throughout this work are tabulated in the Supplementary Information, along with technical details of the microwave apparatus we use. 

For the measurements presented in Fig.~\ref{fig:Theory}, we prepare an equal superposition of $\ket{0}$ and $\ket{1}$ by applying a $\pi / 2$ pulse on the $\ket{0}\leftrightarrow\ket{E}$ transition followed by a $\pi$ pulse on the $\ket{E}\leftrightarrow\ket{1}$ transition. The optical trap is briefly switched off during any microwave pulses in order to avoid varying AC Stark shifts of the transitions across the thermal spatial distribution of molecules. The typical duration for each pulse is $\sim100$\,$\mu$s. We project the phase of the superposition onto the population of the states by reversing this pulse sequence after a hold time $T$ as shown in Fig.\ref{fig:Theory}(d). During $T$ the molecules are confined to the crossed optical dipole trap, and a DC magnetic field is applied in the vertical $z$ direction. With the two microwave frequencies fixed, we observe Ramsey fringes in the form of an oscillating number of molecules in the initial state $\ket{0}$ as a function of~$T$.  

For the measurements presented in Fig.~\ref{fig:ACStark} and Fig.~\ref{fig:BestRamsey}, we find that there is a strong transition from $\ket{0}$ to $(1,4)_4$ just 20\,kHz detuned from the $\ket{E}\leftrightarrow\ket{1}$ transition frequency. We therefore use a modified Ramsey sequence as shown in the inset to Fig.~\ref{fig:BestRamsey} to avoid off-resonantly driving the population out of $\ket{0}$ during the Ramsey pulses.

The frequency of the Ramsey fringes is equal to the two-photon detuning of the microwaves. We fit a model to the fringes (derived in Supplementary Information) which accounts for both two-body collisional loss of molecules and decoherence of the superposition,
\begin{equation}
\begin{split}
    &N(T) = \\
    &\frac{N_i}{2} \left(\frac{1}{1+\frac{T}{T_1}[e-1]}\right)\times\left[e^{-T/T_2}\cos(2\pi(\delta T +\phi))+1\right].
\end{split}
\label{eq:FitFunction}
\end{equation}
Here, $N_i$ is the initial total number of molecules, $T_1$ is the $1/e$ lifetime for molecules in the trap, $T_2$ is the 1/$e$ coherence time, and $\delta$ and $\phi$ are the frequency and phase of the Ramsey fringes. 

To set the magnetic field for a given measurement, we jump the magnetic field to its target value immediately after the STIRAP and then hold 5\,ms before the start of the Ramsey sequence. After the Ramsey sequence is completed, the magnetic field is jumped back to 181.6\,G and held for 5\,ms before the return STIRAP and imaging.

\section{Acknowledgements}
The authors thank M. R. Tarbutt for useful discussions and helpful comments on the early stages of the manuscript. We also thank I. G. Hughes for discussions with regard to the analysis of uncertainties and suggesting the use of the Feldman-Cousins approach. This work was supported by U.K. Engineering and Physical Sciences Research Council (EPSRC) Grants EP/P01058X/1 and EP/P008275/1.

\clearpage

\setcounter{equation}{0}
\setcounter{figure}{0}
\renewcommand{\figurename}{Supp. Fig.}
\renewcommand{\theequation}{S\arabic{equation}}

\section{Supplementary Information}

\section{Molecular constants}
The molecular constants used for the calculation of the rotational and hyperfine structure of $^{87}$Rb$^{133}$Cs are given in Table~\ref{table:Fitting}.

\begin{table}[b]
\begin{tabular*}{\linewidth}{@{\extracolsep{\fill}}ccc}
\hline
\hline
Constant & Value & Ref. \\
\hline
\hline
$B_{v}$ &		490.173~994(45)~MHz & \cite{Gregory2016}\\
$D_{v}$	&	207.3(2)~Hz & \cite{Blackmore2020pccp} \\
$(eQq)_{\text{Rb}}$	&  $-$809.29(1.13)~kHz & \cite{Gregory2016} \\
$(eQq)_{\text{Cs}}$	&	59.98(1.86)~kHz & \cite{Gregory2016}\\
$c_{\text{Rb}}$ & 29.4~Hz & \cite{Aldegunde2008} \\
$c_{\text{Cs}}$ & 196.8~Hz & \cite{Aldegunde2008} \\
$c_{3}$ & 192.4 Hz & \cite{Aldegunde2008} \\
$c_{4}$	&	19.019(105)~kHz & \cite{Gregory2016} \\
$g_{r}$ & 0.0062  &  \cite{Aldegunde2008} \\
$g_{\text{Rb}} \cdot (1-\sigma_{\text{Rb}})$	&	1.8295(24) & \cite{Gregory2016}\\
$g_{\text{Cs}} \cdot (1-\sigma_{\text{Cs}})$	&	0.7331(12) & \cite{Gregory2016}\\
\hline
\hline
\end{tabular*}
\caption{\label{table:Fitting} Constants involved in the molecular Hamiltonian
for $^{87}$Rb$^{133}$Cs. Terms without uncertainties are calculated using
density-functional theory (DFT)~\cite{Aldegunde2008}. Other terms are found by microwave spectroscopy of the rotational transitions~\cite{Gregory2016, Blackmore2020pccp}.}
\end{table}

\section{Crossed optical dipole trap apparatus}

The light for the crossed optical dipole trap (xODT) is generated by a single-mode IPG fibre laser, with wavelength $\lambda=1550~\mathrm{nm}$. The two beams have waists of 81(1)\,$\mu$m and 97(1)\,$\mu$m and cross at an angle of $27^{\circ}$, with both beams propagating in the horizontal plane. There is a frequency difference of 100\,MHz between the beams to avoid interference effects. Both beams are linearly polarised at an angle $\beta$ with respect to the applied magnetic field, which is oriented along the vertical $z$ direction. The angle $\beta$ is set by manually rotating a $\lambda/2$ waveplate in each beam. For measurements with fixed xODT intensity of $15.8$\,kW\,cm$^{-2}$ (Figs.~1 and~3), the trap frequencies experienced by the molecules in the rotational ground state are $(\omega_x, \omega_y, \omega_z)/2\pi=(29(1),119(2),116(2))$\,Hz.

Molecules in different parts of the xODT experience different intensities, with a range determined by the ratio between the beam waist and the width of the molecule sample. The distribution of the molecules is Gaussian with standard deviations $\sigma=\sqrt{k_\mathrm{B}T/m\omega^{2}}$, where \mbox{$T=0.7\,\mu$K} is the temperature of the molecules, such that \mbox{$(\sigma_x, \sigma_y, \sigma_z)\approx(28,6.9,7.0)\,\mu$m.} Due to gravitational sag, the centre of the distribution is $z_{0} = g/\omega_z^{2} \approx 18~\mu$m below the position of peak intensity. Under these conditions, the variation of intensity across the cloud is dominated by the vertical direction and we estimate the $2\sigma$ intensity difference to be
\begin{equation}
\Delta I \approx \frac{8 z_{0} \sigma_{z}}{w_{0}^{2}} I_\mathrm{pk} \approx 0.13 I_\mathrm{pk},
\label{eq: intensity distribution}
\end{equation}
using the mean of the two beam waists, $\omega_0=89\,\mu$m. This represents an upper limit on the intensity variation that could contribute to decoherence in our experiments. 

\section{Microwave apparatus}

To drive the transition between $N=0\leftrightarrow1$, we apply microwaves with a frequency of $2\times B_v \approx 980$\,MHz. The microwaves are generated using a pair of Keysight MXG N5183B 
signal generators, which are synchronised to a common 10\,MHz GPS reference (Jackson Labs Fury). The outputs of both signal generators are connected to a single 3\,W amplifier that drives a homebuilt antenna constructed from 1\,mm diameter copper wire, cut to a length of $\lambda/4\approx7.7$\,cm. The microwaves are polarised such that they drive both $\pi$ and $\sigma^{\pm}$ transitions. Pulses are generated using the built-in pulse modulation mode on the signal generators which are controlled by transistor-transistor logic (TTL) signals derived from a field programmable gate array (FPGA) with microsecond timing resolution. 

In experiments, we find that the resonant frequencies for the transitions depend linearly upon the intensity of the microwaves used to drive the transitions. This is due to off-resonant couplings to other nearby transitions between the rotational states. These energy shifts are $<h\times 3$\,kHz for all measurements shown, and we have tested that the coherence times we measure do not depend upon the intensity of the microwaves used in the state preparation. 

\begin{figure*}[p]
    \centering
    \includegraphics[width=\textwidth]{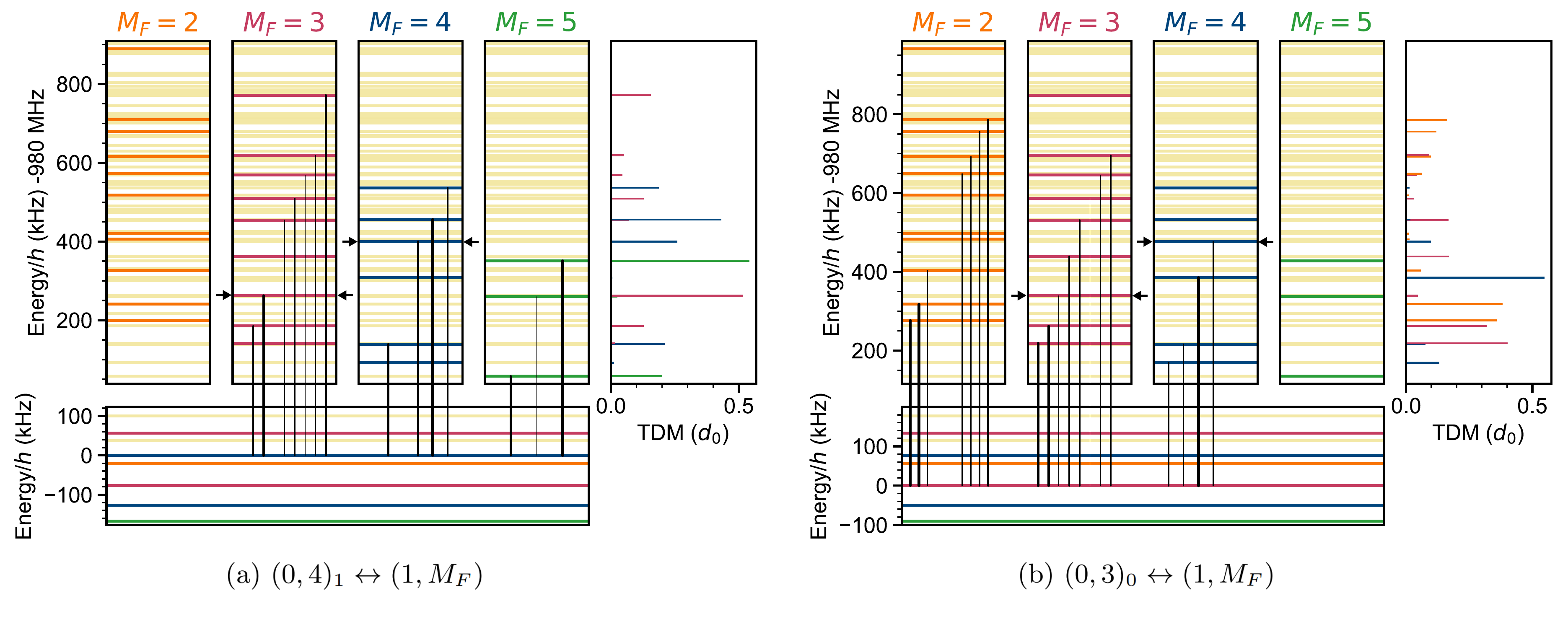}
    \caption{Transitions from the hyperfine states $(0,4)_1$ and $(0,3)_0$ of the $N=0$ rotational state to $N=1$ at 154.5~G. Each state is coloured by the value of $M_F$, allowed transitions have $M_F \rightarrow M_F-1,M_F,M_F+1$ with a strength described by the transition dipole moment (TDM) here shown in units of the permanent dipole moment of the molecule $d_0 \approx 1.23~\mathrm{D}$. The vertical lines connect the labelled initial state for (a) and (b) to excited states via the electric dipole allowed transitions, the thickness of the line corresponds to the strength of the transition. For both (a) and (b) the initial state is labelled as the zero of energy. The target states used in this work are indicated by the arrows. }
    \label{fig:n0transitions}
\end{figure*}

\begin{figure*}[p]
    \centering
    \includegraphics[width=\textwidth]{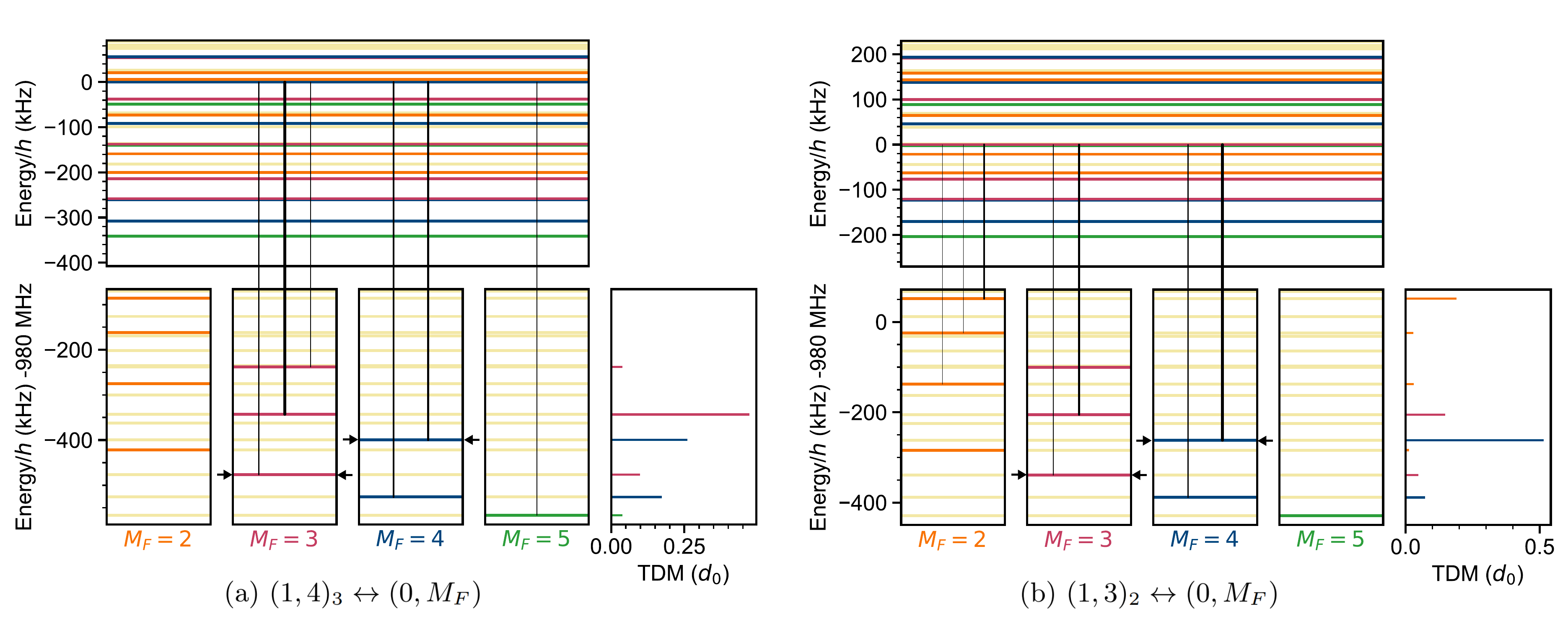}
    \caption{Transitions from the hyperfine states $(1,4)_3$ and $(1,3)_2$ of the $N=1$ rotational state to $N=0$ at 154.5~G. Each state is coloured by the value of $M_F$, allowed transitions have $M_F \rightarrow M_F-1,M_F,M_F+1$ with a strength described by the transition dipole moment (TDM) here shown in units of the permanent dipole moment of the molecule $d_0 \approx 1.23~\mathrm{D}$. The vertical lines connect the labelled initial state for (a) and (b) to the hyperfine sub-levels of the ground state via the electric dipole allowed transitions, the thickness of the line corresponds to the strength of the transition. For both (a) and (b) the initial state is labelled as the zero of energy. The target states used in this work are indicated by the arrows. }
    \label{fig:n1transitions}
\end{figure*}

In Supp. Fig. \ref{fig:n0transitions} and Supp. Fig. \ref{fig:n1transitions} we show the available transitions for the two ground states used at 154.5~G [$(0,4)_1$ and $(0,3)_0$] as well as the available transitions from the excited states $(1,4)_3$ and $(1,3)_2$ at a magnetic field of 154.5~G. For a given sub-level of $N=1$ there are fewer allowed transitions back to $N=0$ due to the smaller total number of hyperfine sub-levels and so off-resonant coupling is less of a concern. We note that any $N=1$ component which remains during the Ramsey hold will quickly dephase (coherence time less than 1\,ms), and so if present would simply contribute a non-zero molecule number background to the Ramsey fringes. The absence of any non-zero background signal in Fig.~3 indicates that this is not an issue.

\section{States and transitions used in this work}
The molecular states that we label by $(N,M_F)_k$ in the main text are a superposition of the products of different molecular rotational states and nuclear spin states. 
To determine the coefficients of these states, we construct the Hamiltonian in the fully uncoupled basis with basis functions $\ket{N,M_N,m_\mathrm{Rb},m_\mathrm{Cs}}$. The quantum number $M_F=M_N+m_\mathrm{Rb}+m_\mathrm{Cs}$ is conserved when the laser polarisation is parallel to the magnetic field ($\beta=0$), but not otherwise. For each eigenstate, we calculate the expectation value $\bra{\psi}F_z\ket{\psi}$ and label the state with the nearest integer value of $M_F$. We then order the states by energy to determine $k$.
The composition of each of the states used in this work is shown in table~\ref{table:StateCompositions}, with coefficients rounded to 1 part in $10^3$. This rounding causes the table to omit coefficients that are non-zero and on the order of 1 part in 10$^5$ to 1 part in 10$^6$ for basis states with $N=2$ in the ground rotational state. There are coefficients with a similar magnitude for basis states with $N=3$ in the first rotationally excited state.

\section{Derivation of equation 1 and Magnetic field limit on the coherence time at $B=154.52$\,G}

\begin{figure}[p]
    \centering
    \includegraphics[width=0.46\textwidth]{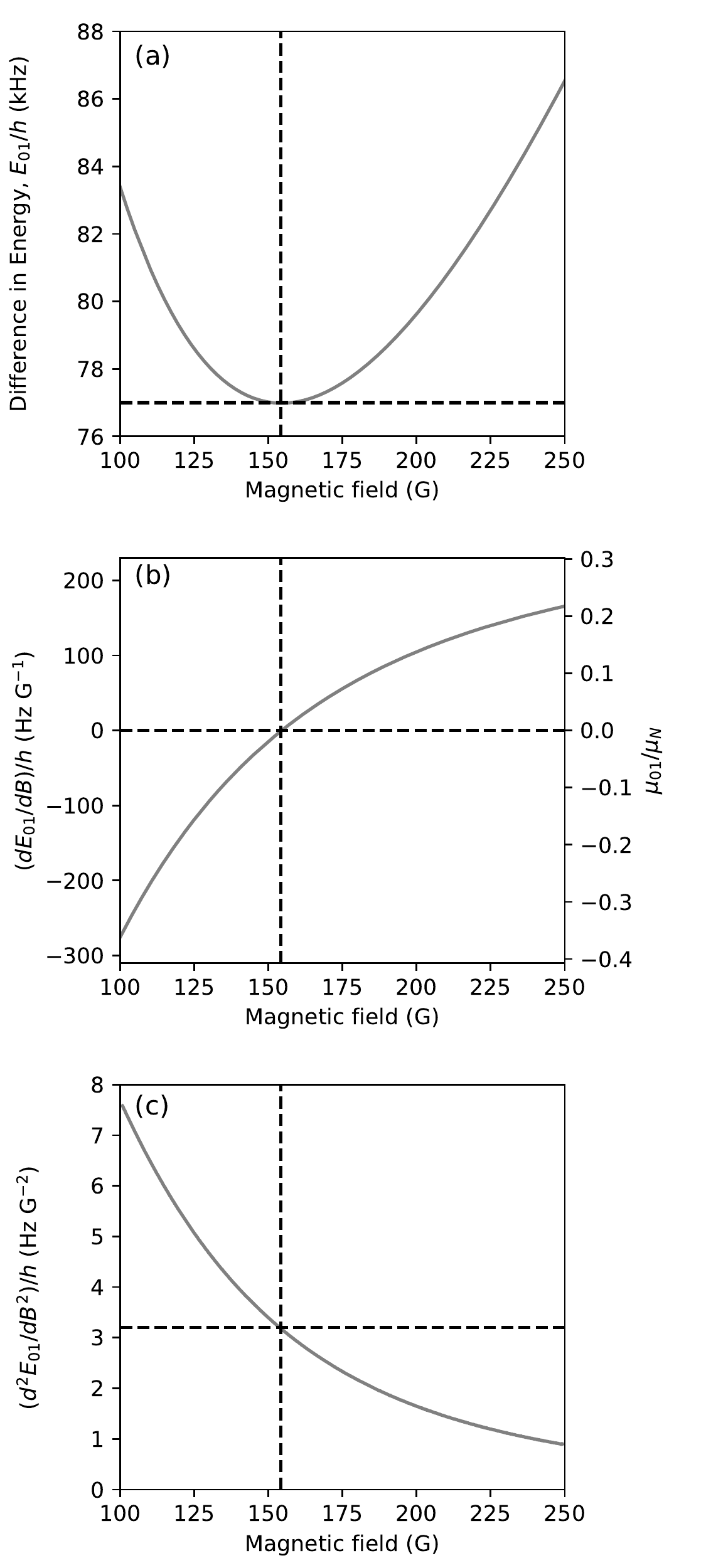}
    \caption{(a) Differential Zeeman shift between the states $E_{01} = |E_{\ket{0}} - E_{\ket{1}}|$ along with the (b) first and (c) second derivatives of $E_{01}$ with respect to magnetic field. The first derivative $dE_{01}/dB = \mu_{01}=0$ at a magnetic field of $B=154.52$\,G indicating a turning point in $E_\mathrm{01}$.}
    \label{fig:EnergyDerivatives}
\end{figure}

\begin{figure}[p]
    \centering
    \includegraphics[width=0.46\textwidth]{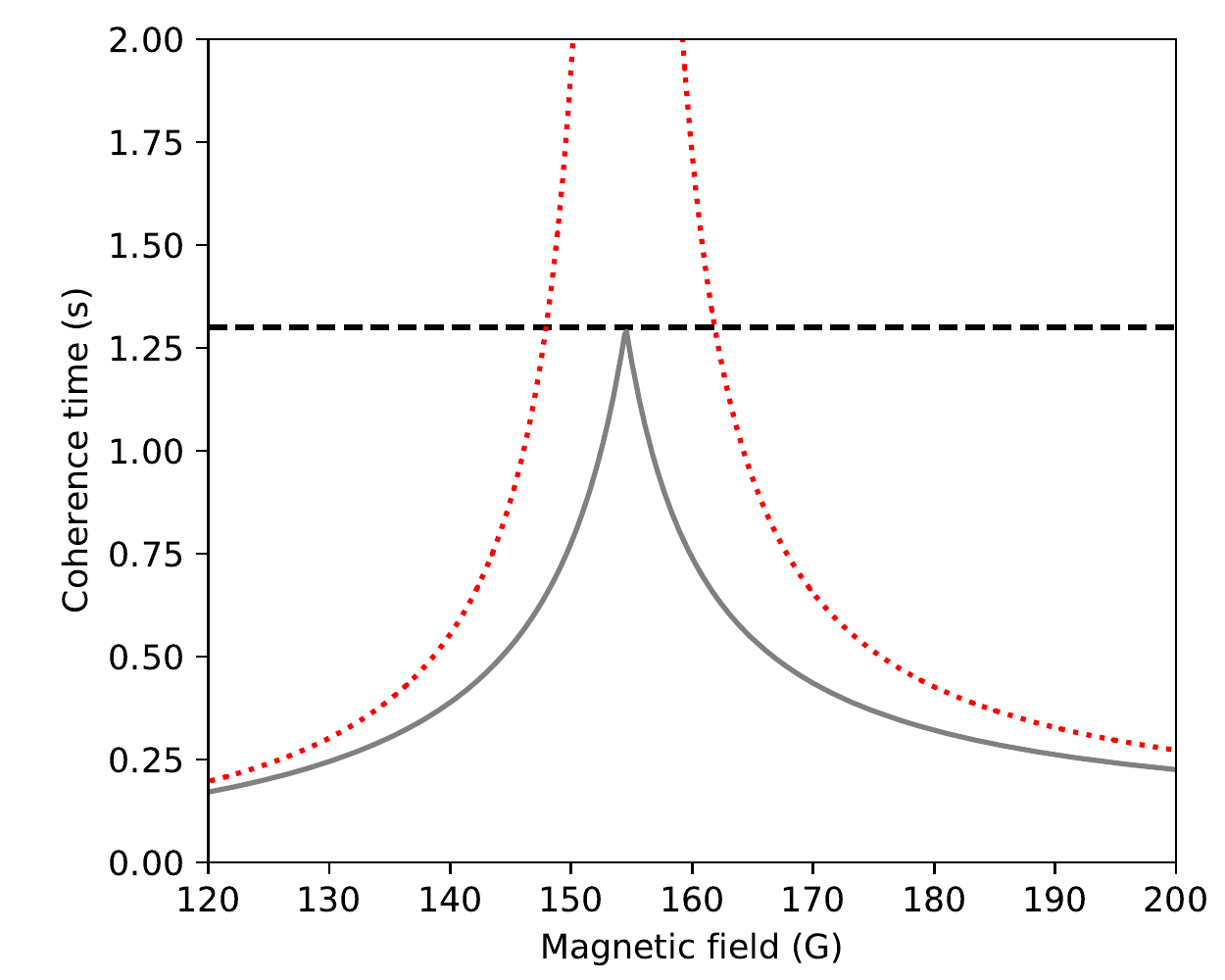}
    \caption{Coherence time calculated for $\Delta B=35$\,mG. The red dotted line indicates the expected coherence time for magnetic field noise alone, calculated using the expression in Eq.~\ref{eq:MagneticFieldDecoherenceOnly}. The gray solid line is the complete model including decoherence from the tensor light shifts ($T^*_2=1.3$\,s) that is present when the trap laser is polarised parallel to the magnetic field direction $(\beta=0)$, as presented in the main text. The horizontal dashed line indicates the coherence time limited by the tensor light shifts alone, which dominates the coherence time for magnetic fields across a broad range of magnetic fields around the turning point. 
    }
    \label{fig:TensorLightShiftsDominate}
\end{figure}

The coherence time $T_2$ is limited by variation $\Delta E_{01}$ in the energy difference between the two states such that
\begin{equation}
T_2 = \frac{h}{|\Delta E_{01}|},
\label{eq:GeneralEnergyRelation}
\end{equation}
where $h$ is the Planck constant. The energy difference at a given magnetic field
\begin{equation}
E_{01}(B) =  |E_{\ket{0}}(B) - E_{\ket{1}}(B)|
\end{equation}
can be calculated from the energies of the two states, $E_{\ket{0}}(B)$ and $E_{\ket{1}}(B)$. We plot $E_{01}$ as a function of magnetic field in Supp.~Fig.~\ref{fig:EnergyDerivatives}(a). The magnitude of $\Delta E_{01}$ is the difference between the maximum and minimum value of $E_\mathrm{01}$ experienced in a given measurement. 

To explain the results shown in Fig.\,1, we must evaluate $E_\mathrm{01}$ across the range of magnetic fields defined by the magnetic field noise $\Delta B$. Away from the turning point at $B_0=154.52$\,G, the minimum and maximum values of $E_{01}$ are found at $B \pm \Delta B / 2$. The variation in energy can therefore be evaluated by
\begin{equation}
\Delta E_{01}=|E_{01}(B+\Delta B / 2)- E_{01}(B-\Delta B / 2)|.
\label{eq:FarFromTurningPointVariation}
\end{equation}
At the turning point, the minimum and maximum values of $E_{01}$ are found at $B_0$ and $B_0 \pm \Delta B / 2$ respectively. The variation in energy here is
\begin{equation}
\Delta E_{01}=| E_{01}(B_0\pm\Delta B / 2)- E_{01}(B_0) |.
\label{eq:TurningPointVariation}
\end{equation}
The transition between these two regimes occurs when $|B-B_0|\approx\Delta B$. 

\subsection{Derivation of Equation 1} 

When the trap laser is polarised parallel to the magnetic field direction $(\beta=0)$, we find that the decoherence is dominated by the tensor light shifts across a wide range of magnetic fields around the turning point (see Supp.~Fig.~\ref{fig:TensorLightShiftsDominate}). As such we can reasonably approximate the magnetic field variation using just Eq.~\ref{eq:FarFromTurningPointVariation}.
To arrive at the fit function presented in Eq.~1, we calculate the Taylor expansion of Eq.~\ref{eq:FarFromTurningPointVariation} to find,
\begin{equation}
\Delta E_{01} = \frac{dE_{01}}{dB} \left(\Delta B\right) + \frac{1}{2}\frac{d^2E_{01}}{dB^2}\left(\Delta B\right)^2 + ...
\label{eq:TaylorExpansion1}
\end{equation}
The first and second derivatives of energy with respect to magnetic field are plotted in Supp.~Fig.~\ref{fig:EnergyDerivatives}(b) and (c), respectively. The second derivative of $E_{01}$ is two orders of magnitude smaller than the first derivative at magnetic fields where the tensor light shifts do not dominate. For small variations in magnetic field $\Delta B < 1$\,G, we can therefore approximate
\begin{equation}
\Delta E_{01} \approx \frac{dE_{01}}{dB}(\Delta B) \equiv \mu_{01}\Delta B,
\label{eq:ApproxVariation}
\end{equation}
using only the first term in Eq.~\ref{eq:TaylorExpansion1}.
Substituting Eq.~\ref{eq:ApproxVariation} into Eq.~\ref{eq:GeneralEnergyRelation} we find the coherence time limited by magnetic field noise
\begin{equation}
T'_2 \approx \frac{h}{|\mu_{01}|\Delta B}.
\label{eq:MagneticFieldDecoherenceOnly}
\end{equation}
To include the differential tensor light shifts as an additional source of decoherence, with coherence time $T^*_2$, we combine the coherence times as
\begin{equation}
\frac{1}{T_2} = \frac{1}{T'_2} + \frac{1}{T^*_2},
\end{equation}
to find the fit function
\begin{equation}
T_2 = \left(\frac{|\mu_{01}|~\Delta B}{h} + \frac{1}{T^*_2}\right)^{-1}.
\end{equation}

\subsection{Magnetic field limit on the coherence time at $B=154.52$\,G}

We can estimate the limit placed on the coherence time by $\Delta B$ by at the turning point by performing a Taylor expansion of Eq.~\ref{eq:TurningPointVariation}
\begin{equation}
\Delta E_{01} = \frac{dE_{01}}{dB} \left(\frac{\Delta B}{2}\right) + \frac{1}{2}\frac{d^2E_{01}}{dB^2}\left(\frac{\Delta B}{2}\right)^2 + ...
\label{eq:TaylorExpansion2}
\end{equation}
At this magnetic field, $dE_{01}/dB=0$, and so only the second derivative contributes. We calculate $d^2E_{01}/dB^2=h\times3.2$\,Hz\,G$^{-2}$ (see Supp.~Fig.~\ref{fig:EnergyDerivatives}(c)). For $\Delta B=35$\,mG, we therefore find
\begin{equation}
\Delta E_{01} = \frac{1}{2} \times (h\times3.2) \times \left(\frac{0.035}{2}\right)^{2} = h\times0.49\mathrm{\,mHz},
\end{equation}
with a corresponding coherence time of 
\begin{equation}
T'_2 = \frac{1}{0.49\mathrm{\,mHz}} = 2.0\times10^3  \mathrm{\,s}.
\end{equation}
This is remarkably long and validates the omission of the second order term in Eq.~S11 from the fit function in Eq.~1.

\section{Derivation of equation 11}

The function used to fit the observed Ramsey fringes is given in Eq.~11. 

In the absence of decoherence and collisional loss of molecules from the trap, the Ramsey fringes are described by
\begin{equation}
N(T) = \frac{N_i}{2}\left[\cos(2\pi(\delta T + \phi)) + 1\right],
\end{equation}
where $N(T)$ is the number of molecules remaining in state $\ket{0}$, $N_i$ is the total number of molecules, $T$ is the hold time between the Ramsey pulses, and $\delta$ and $\phi$ are the frequency and phase of the Ramsey fringes. We include decoherence, with a characteristic $1/e$ coherence time $T_2$ as
\begin{equation}
N(T) = \frac{N_i}{2}\left[e^{-T/T_2}\cos(2\pi(\delta T + \phi)) + 1\right],
\end{equation}
where the addition of the exponential term reduces the contrast of the Ramsey fringes as $T$ increases. 

Collisional loss of molecules from the trap reduces the total number of molecules remaining in the sample, but does not affect the contrast of the fringes. We have previously shown that these losses are due to fast optical excitation of two-body collision complexes by the trap light~\cite{Gregory2020}, and the rate limiting step for this loss mechanism is therefore two-body~\cite{Gregory2019}. Accordingly, the rate equation for the density of ground-state molecules $n(t)$ is
\begin{equation}
\frac{dn}{dt} = -K_2 n(t)^{2},
\end{equation}
where $K_2$ is a two-body rate coefficient which characterises the loss with units $\mathrm{m}^{3}\,\mathrm{s}^{-1}$. This equation can be rewritten in terms of the molecule number $N(t)$ by introducing an effective volume $V_{\textrm{eff}}= (m\bar{\omega}/(4\pi k_{\textrm{B}}T_{\textrm{m}})^{-(3/2)}$ which depends on the temperature of the molecules $T_{\textrm{m}}$ and the geometric mean of the trap frequencies $\bar{\omega}=(\omega_x\omega_y\omega_z)^{1/3}$. This yields
\begin{equation}
\frac{dN}{dt} = -\frac{K_2}{V_{\textrm{eff}}} N(t)^{2}.
\end{equation}
To simplify the solution of this equation we assume that the temperature remains constant throughout the measurement. Rearranging and integrating then leads to
\begin{equation}
N(t) = \frac{N_i}{1+\frac{K_2}{V_{\textrm{eff}}} N_i t},
\end{equation}
where $N_i$ is the initial molecule number. To find the $1/e$ time which characterises this loss $T_1$ we must evaluate
\begin{equation}
N(T_1) = \frac{N_i}{1+\frac{K_2}{V_{\textrm{eff}}} N_i T_1}=\frac{N_i}{e},
\end{equation}
which by rearrangement leads to
\begin{equation}
\frac{K_2}{V_{\textrm{eff}}} N_i = \frac{e-1}{T_1}.
\end{equation}
Substituting Eq.~S15 back into Eq.~S13 yields
\begin{equation}
N(t) = \frac{N_i}{1+\frac{t}{T_1}(e-1)}.
\end{equation}

To describe the Ramsey fringes in the presence of both decoherence and collisional loss, we must replace $N_i$ in Eq.\ S10 with the expression for $N(t)$ in Eq.\ S16 to find the fit function given in Eq.\ 11 as a function of the Ramsey time $T$
\begin{equation}
\begin{split}
    &N(T) = \\
    &\frac{N_i}{2} \left(\frac{1}{1+\frac{T}{T_1}[e-1]}\right)\times\left[e^{-T/T_2}\cos(2\pi(\delta T +\phi))+1\right].
\end{split}
\end{equation}
It is worth noting that whilst $T_1$ is the time for the molecule number to fall to $1/e$ of the initial value, the decay is not exponential and so waiting $2T_1$ does not lead to the molecule number falling to $1/e^2$ of the initial value. This is an artefact of the density-dependent character of the two-body loss.

\section{Matrix elements for $H_\mathrm{quad}$ and $H_\mathrm{AC}$}

The dominant terms that contribute to the differential AC Stark shift between hyperfine states are $H_\mathrm{quad}$ and $H_\mathrm{AC}$. To be explicit, and to demonstrate the off-diagonality in $N$, we give the matrix elements for each of these terms here. 
\begin{equation}
\begin{split}
    &\bra{N,M_N,m_\mathrm{Rb},m_\mathrm{Cs}}H_\mathrm{quad}\ket{N',M_N',m_\mathrm{Rb}',m_\mathrm{Cs}'}=\\
    &\sum_{M=-2}^2\bigg\{\sqrt{(2N+1)(2N'+1)}(-1)^M\\
    &\times
    \begin{pmatrix}
    N & 2 & N'\\
    -M_N & M & M_N'
    \end{pmatrix}
    \begin{pmatrix}
    N & 2 & N'\\
    0 & 0 &0
    \end{pmatrix}\\
    &\times\bigg[\left(\frac{(eqQ)_\mathrm{Rb}}{4}\right)(-1)^{M_N+I_\mathrm{Rb}-m_\mathrm{Rb}}\\
    &\times
    \frac{
    \begin{pmatrix}
    I_\mathrm{Rb} & 2 & I_\mathrm{Rb}\\
    -m_\mathrm{Rb} & -M & m_\mathrm{Rb}'
    \end{pmatrix}}
    {\begin{pmatrix}
    I_\mathrm{Rb} & 2 & I_\mathrm{Rb}\\
    -I_\mathrm{Rb} & 0 & I_\mathrm{Rb}
    \end{pmatrix}}\delta_{m_\mathrm{Cs},m_\mathrm{Cs}'}\\
    &+\left(\frac{(eqQ)_\mathrm{Cs}}{4}\right)(-1)^{M_N+I_\mathrm{Cs}-m_\mathrm{Cs}}\\&\times
    \frac{
    \begin{pmatrix}
    I_\mathrm{Cs} & 2 & I_\mathrm{Cs}\\
    -m_\mathrm{Cs} & -M & m_\mathrm{Cs}'
    \end{pmatrix}}
    {\begin{pmatrix}
    I_\mathrm{Cs} & 2 & I_\mathrm{Cs}\\
    -I_\mathrm{Cs} & 0 & I_\mathrm{Cs}
    \end{pmatrix}}\delta_{m_\mathrm{Rb},m_\mathrm{Rb}'}\bigg]\bigg\},
\end{split}\label{eq:Hquad_explicit}
\end{equation}
In the above, terms in parentheses are Wigner-3j symbols, $\delta_{A,B}$ represents the Kronecker delta function and the coefficients have the same definition as in~\cite{Gregory2016}.
\begin{equation}
\begin{split}
&\bra{N,M_N}H_\mathrm{AC}\ket{N',M_N'} = -\frac{I\alpha^{(0)}}{2\epsilon_0c}\delta_{N,N'}\delta_{M_N,M_N'}\\
-&\frac{I \alpha^{(2)}}{2\epsilon_0c} \sum_{M} d_{M 0}^{2}(\beta)(-1)^{M_{N}^{\prime}} \sqrt{(2 N+1)\left(2 N^{\prime}+1\right)}\\
\times&\left(\begin{array}{ccc}{N^{\prime}} & {2} & {N} \\ {0} & {0} & {0}\end{array}\right)\left(\begin{array}{ccc}{N^{\prime}} & {2} & {N} \\ {-M_{N}^{\prime}} & {M} & {M_{N}}\end{array}\right).
\label{eq:ACStark_elements}
\end{split}
\end{equation}
Here, $I$ is the laser intensity, $\delta_{A,B}$ is a Kronecker delta and $d^2_{M0}(\beta)$ is a reduced Wigner rotation matrix. The term proportional to the isotropic part, $\alpha^{(0)}$, produces an equal energy shift of all $(N,M_N)$. The term proportional to the anisotropic part, $\alpha^{(2)}$, has more complicated behavior: for $N>0$ it has elements both diagonal and off-diagonal in $N, M_N$ that depend on $\beta$.

\section{Validating the form of Equation~4}

\begin{figure}[p]
    \centering
    \includegraphics[width=0.46\textwidth]{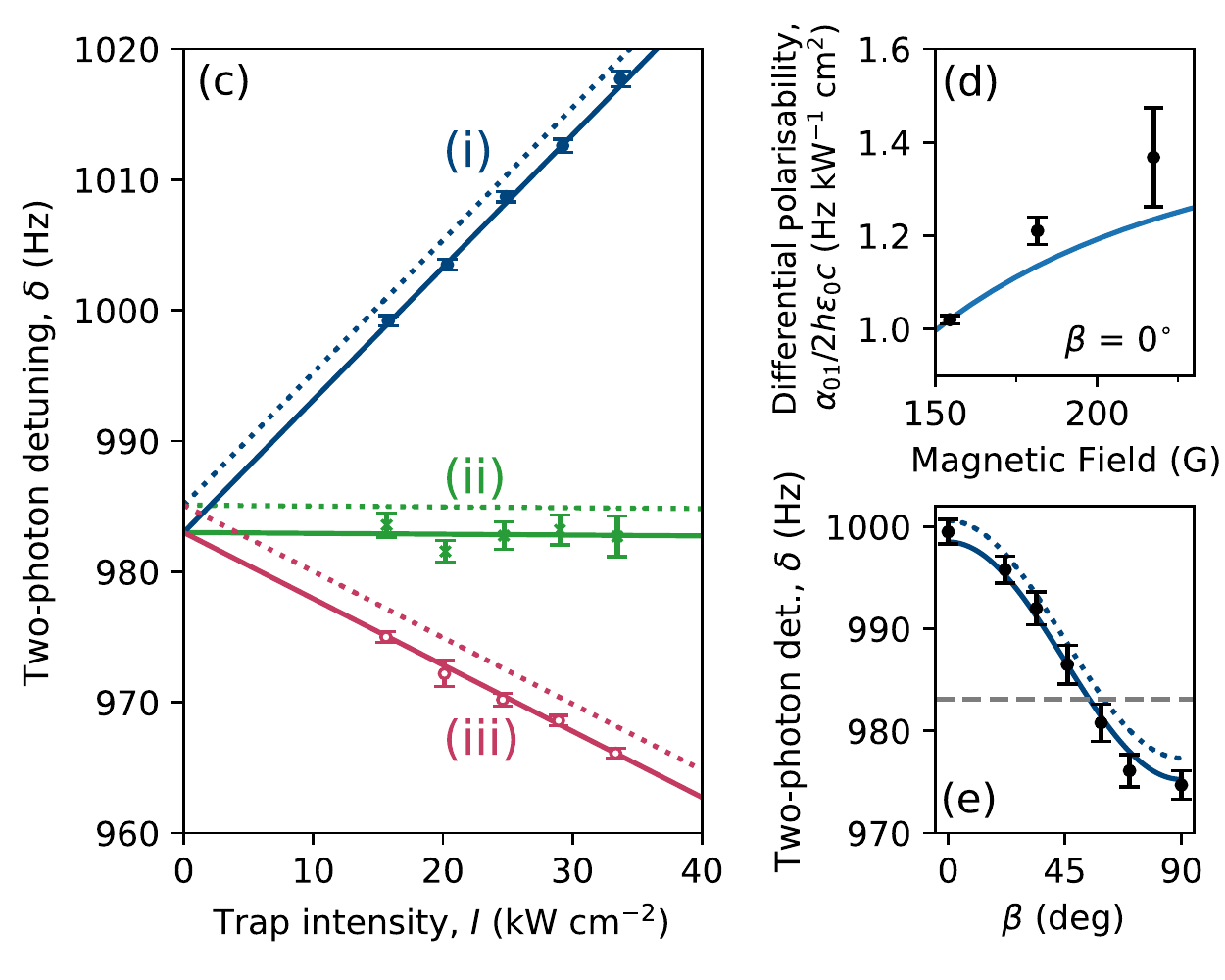}
    \caption{Differential tensor light shifts from the full rotation and hyperfine calculation. Here we have reproduced the experimental results shown in Fig.\,2(c)-(e), together with the results of the simple fit function given by Eq.(3) and Eq.(4). In addition, the dotted lines indicate the output of the full rotation and hyperfine calculation, including the difference in the free-space detuning highlighted in the main text. The solid lines have the free-space detuning fixed to be $\delta_0=983.0$\,Hz. We see that the intensity dependence is nearly identical to the simpler model presented in Fig.\,2(c) and (e), which indicates that our simpler equations capture the behaviour of the differential tensor light shifts well.  
    }
    \label{fig:FullTheory}
\end{figure}

Our model for the rotational and hyperfine structure of RbCs is able to replicate the structure of the AC Stark shift observed in experiments, and is used to calculate the magnetic field dependencies presented in Fig.~1 and Fig.~2(d). However, for Fig. 2(c) and (e) a simplified fit function is used, given by Eq.~2 and Eq.~3, and we use our full Hamiltonian to calculate the numerical factor $X(B)$. The simpler fit function was then used to find the optimal value for the anisotropic polarisability $\alpha^{(2)}$ and the free-space detuning $\delta_0$. In Supp. Fig.~\ref{fig:FullTheory} we show the calculations using the full Hamiltonian for all of the results in Fig.~2(c)-(e). We see that our full model is well described by the simpler fit function we use in the main text. 

\section{Additional systematic uncertainties in the measurement of $\alpha^{(2)}$}
There are additional systematic contributions to the uncertainty in $\alpha^{(2)}$. Uncertainty in $I$ contributes additional uncertainty of $\pm2\,\%$ to $\alpha^{(2)}$. There is also uncertainty from the compositions of the states which is more difficult to quantify due to the large number of parameters in the Hamiltonian. The largest contribution to the mixing of $N$ is from the Rb electric quadrupole coupling, characterised by the constant $(eQq)_\mathrm{Rb}$; the uncertainty from this parameter contributes an uncertainty in $\alpha^{(2)}$ of $\pm1\,\%$. The fitted value of $\alpha^{(2)}$ lies intermediate between the two values we previously obtained from microwave spectra on the $N=0\leftrightarrow1$ transitions~\cite{Gregory2017}.

\section{Variation of the AC Stark effect with magnetic field}
\begin{figure}[t]
    \centering
    \includegraphics[width=0.46\textwidth]{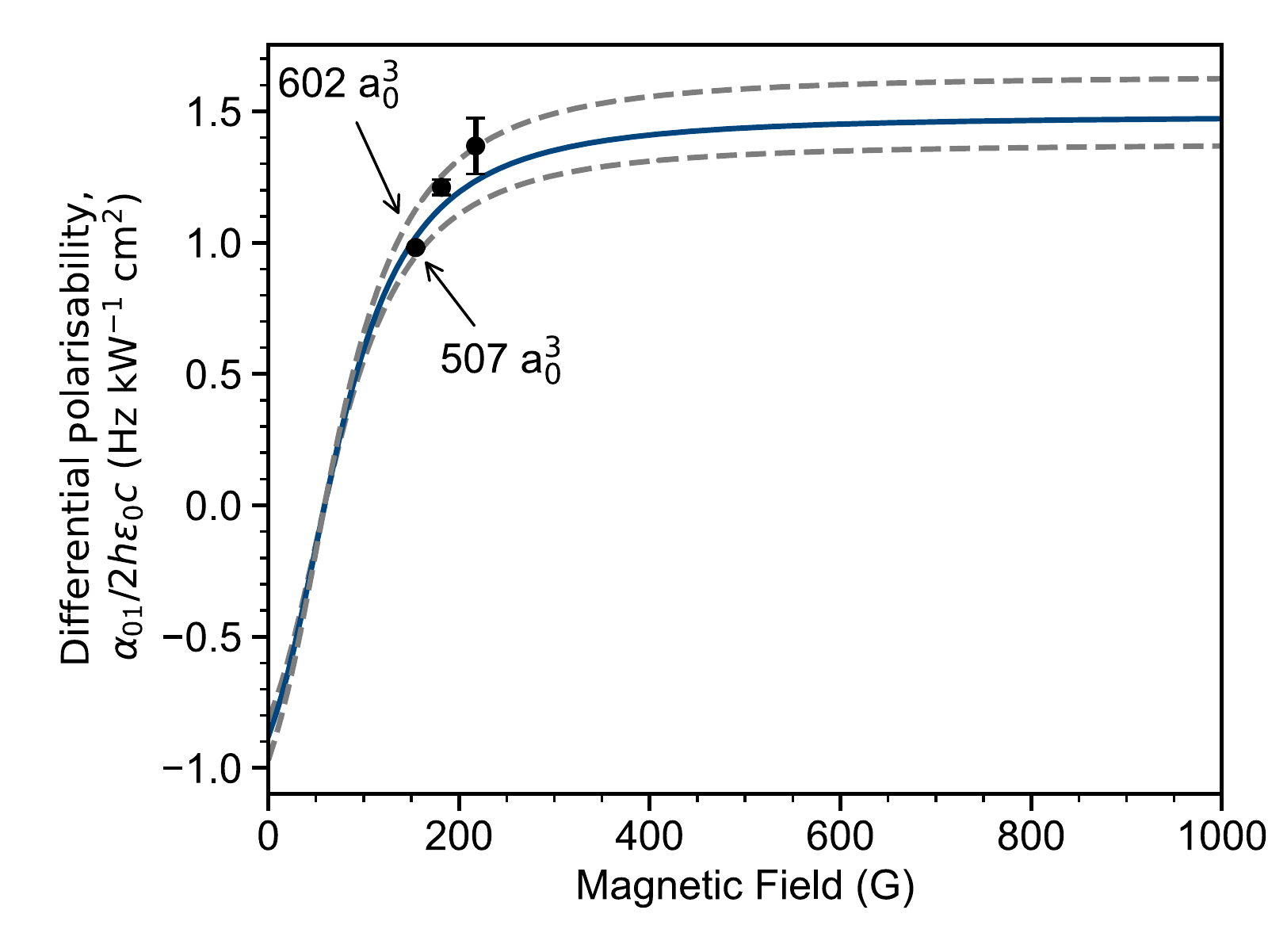}
    \caption{
    The gradient of the differential AC Stark shift as a function of the applied magnetic field. The data points and the solid blue line are the same as in Fig.~2(d). The gray dashed lines correspond to the two values of $\alpha^{(2)}$ reported in~\cite{Gregory2017} and are labelled by $\alpha^{(2)}/4\pi\epsilon_0$.
    }
    \label{fig:LargeMagnetic}
\end{figure}
The differential AC Stark shift is highly dependent on the nuclear spin state of the molecules. In the experiment we investigate only a small range of magnetic fields. Here we present the full calculation as a function of magnetic field from 0 to 1000~G.

To determine this behaviour we extract the eigenvalues of the Hamiltonian as a function of the strength of the applied magnetic field for zero intensity. Each continuous energy level is labelled by $M_F$ which is determined by the expectation value $\bra{\psi}F_z\ket{\psi}$ for each eigenstate $\ket{\psi}$. We repeat this analysis for a second intensity $I=60~\mathrm{kW\,cm^{-2}}$ with $\beta=0^{\circ}$. As the differential AC Stark shift we are investigating is linear, we extract the differential polarisability by determining the slope of the change of the energy difference between the two states as a function of intensity. 

In Supp.~Fig.~\ref{fig:LargeMagnetic} we show this gradient as a function of the applied magnetic field. As the magnetic field increases the gradient of the AC Stark shift eventually reaches some asymptotic value. This occurs because there are multiple competing terms in the total Hamiltonian that mix both rotational and nuclear spin states. However the Zeeman effect acts mostly on the nuclear spin states, which ultimately causes the nuclear spin states to decouple such that the overall molecular wavefunction is best represented by the product of a single rotational state $\ket{N,M_N}$ and nuclear spin state $\ket{m_\mathrm{Rb},m_\mathrm{Cs}}$. At high magnetic field $\ket{0}$ and $\ket{2}$ differ by 1 in the value of $m_\mathrm{Rb}$ and by 2 in the value of $m_\mathrm{Cs}$; as the dominant off-diagonal term is the Rb nuclear electric quadrupole coupling, we believe that it is the difference in Rb nuclear spin projection that contributes to the large differential effective polarisability.

\section{Estimating the minimum coherence time in Fig. 4}

\begin{figure*}[t]
    \centering
    \includegraphics[width=\textwidth]{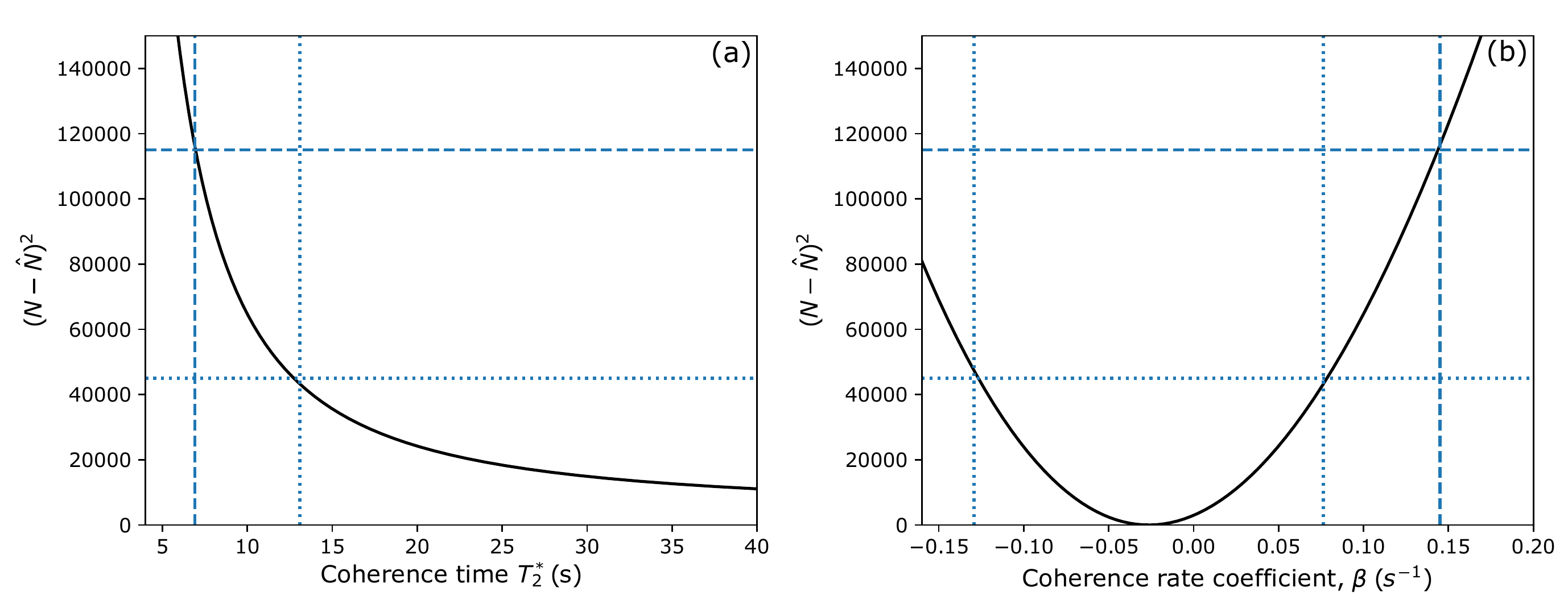}
    \caption{Change in the residual sum of squares as a function of the free parameters (a) $T^*_2$ and (b) $\beta=1/T^*_2$ when fitting the results presented in Fig.~3. The minimum value of the residuals is 8604837, found at the minima seen in (b). This value is subtracted from the $y$-axes of both sets of data to give the change in residuals from this minimum value. The dotted lines indicate $\pm1\sigma$ from the mean value found in (b). The dashed lines indicate the 90\,\% confidence interval for the minimum value of the coherence time $T_2$ consistent with our results. 
    }
    \label{fig:ErrorSurfaces}
\end{figure*}

\begin{figure}[t]
    \centering
    \includegraphics[width=0.46\textwidth]{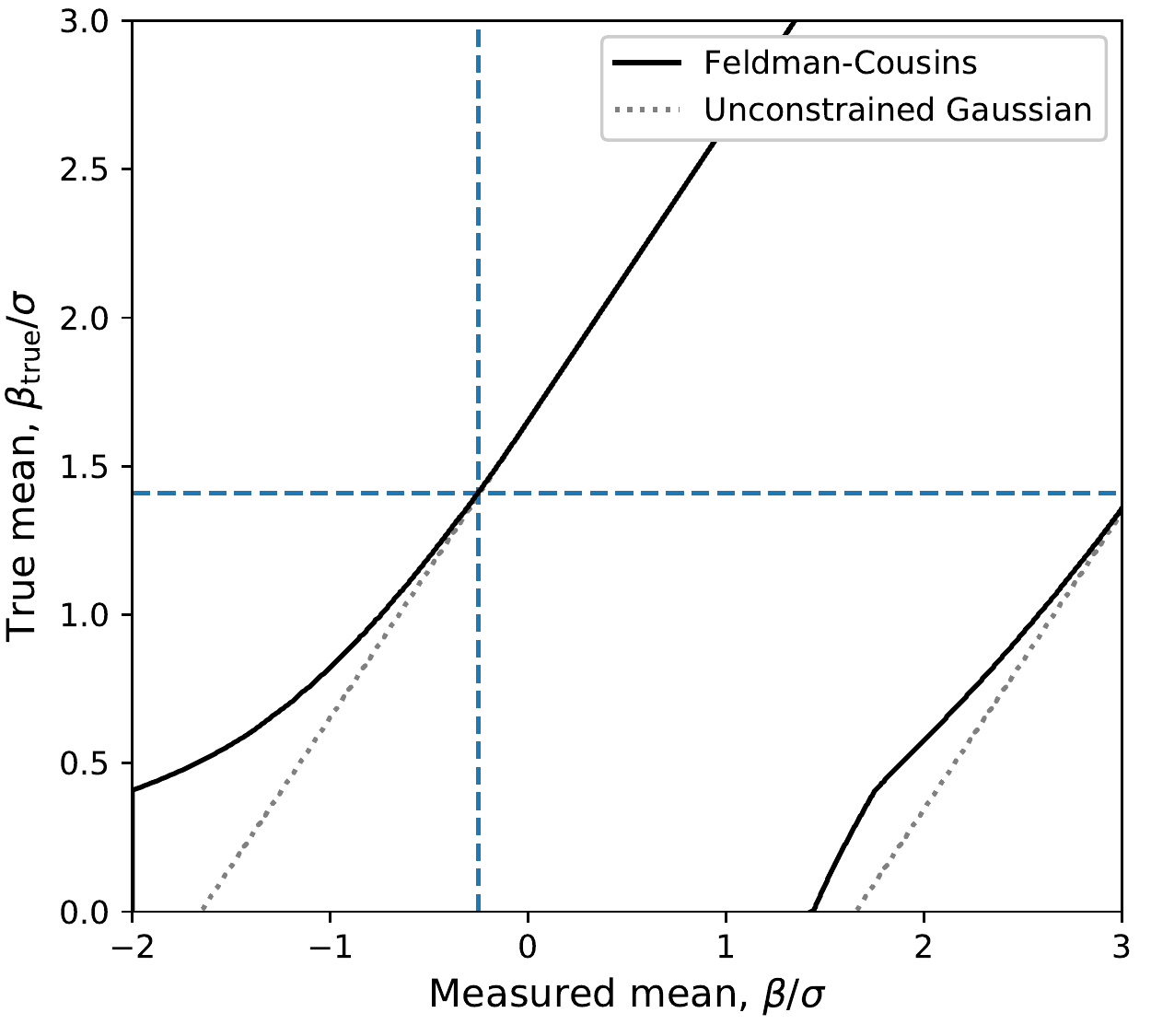}
    \caption{The 90\,\% confidence bands for the mean of a Gaussian distribution which is constrained to be non-negative using the Feldman-Cousins approach (solid lines). The 90\,\% confidence bands for an unconstrained Gaussian distribution are also shown (dotted lines). The dashed vertical lines indicate the measured value of $\beta/\sigma=-0.251$. The dashed horizontal line shows where the vertical line intersects the 90\,\% upper boundary when $\beta_\mathrm{true}\approx1.41\sigma$.  
    }
    \label{fig:ConfidenceIntervals}
\end{figure}

Our Ramsey measurements yield oscillations in the number of molecules remaining in $\ket{0}$ as a function of time. For the results in Figs.~1 and 2, we fit the oscillations using the model shown in Eq.~11. Fitting is performed by a least-squares regression. 
$T^*_2$ cannot be determined from the results shown in Fig. 3, because the sum of squares of residuals decreases continuously as $T_2\rightarrow\infty$, as shown in Supp.~Fig.~\ref{fig:ErrorSurfaces}(a).

To estimate the minimum value of $T^*_2$ which is consistent with our results in Fig.~3, we replace the fit parameter with a coherence rate coefficient $\beta=1/T_2$ such that the fit function becomes, 
\begin{equation}
\begin{split}
    &N(T) = \\
    &N_i \left(\frac{1}{1+\frac{T}{T_1}\times[e-1]}\right)\times\frac{1}{2}\times\left[e^{-\beta T}\cos(\delta T +\phi)+1\right].
\end{split}
\label{eq:FitFunction}
\end{equation}
As $T_2\rightarrow\infty$, $\beta\rightarrow 0$, and we find a minimum in the residual sum of squared as shown in Supp.~Fig.~\ref{fig:ErrorSurfaces}(b). We find an optimum value of $\beta=-0.026$\,s$^{-1}$ with root mean square (RMS) deviation $\sigma=0.103$\,s$^{-1}$. 

In the main text we report a 90\,\% confidence interval for the minimum value of $T^*_2$, which we find using the methods laid out by Feldman and Cousins~\cite{Feldman1998}; this approach unifies the treatment of one- and two-sided confidence intervals. We apply the method to a model with Gaussian statistics
\begin{equation}
P(\beta|\beta_\mathrm{true})=\frac{1}{\sigma\sqrt{2\pi}}\exp\left(-\frac{(\beta-\beta_\mathrm{true})^{2}}{2\sigma^{2}}\right),
\end{equation}
where $\beta$ is the measured value of $\beta_\mathrm{true}$, the true value of the parameter, with RMS deviation~$\sigma$. We consider the case where $\beta_\mathrm{true}>0$; this is valid for our experiments as a negative value of $\beta$ is non-physical, corresponding to increasing coherence over the experiment. The 90\,\% confidence intervals are shown in Supp. Fig.~\ref{fig:ConfidenceIntervals}, which we calculate using the same procedure as described in~\cite{Baron2017}. The vertical line at $\beta/\sigma\approx-0.026/0.103\approx-0.251$ indicates our measured value, which puts a corresponding 90\,\% upper limit on $\beta_\mathrm{true}\approx1.41\sigma\approx0.145$\,s$^{-1}$. This upper limit on $\beta$ corresponds to a lower limit on $T^*_2=1/\beta_\mathrm{true}=6.9$\,s (90\,\% confidence level). 

\section{Confirming the absence of collisional shifts}
\begin{figure*}[t]
    \centering
    \includegraphics[width=\textwidth]{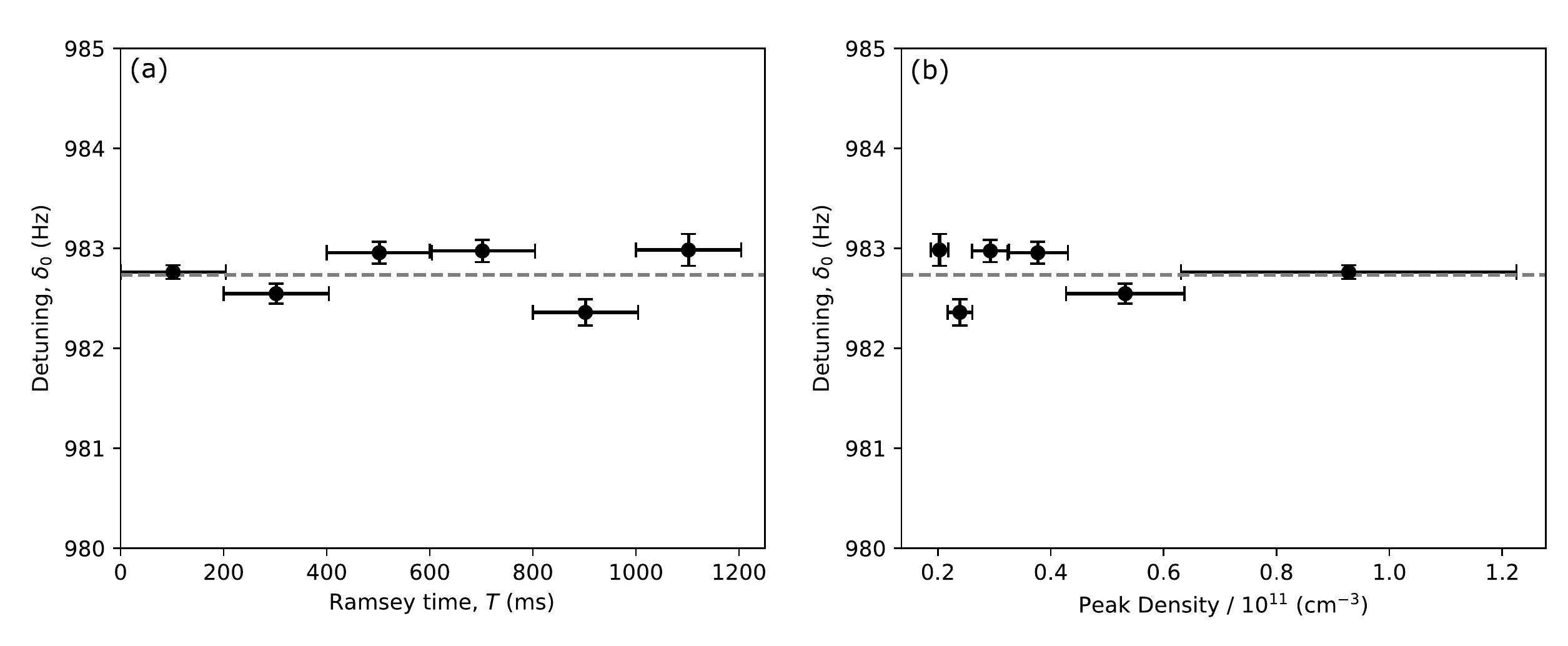}
    \caption{Detuning measured from the Ramsey fringes presented in Fig.~3, but segmented into shorter 204\,ms time intervals. As the Ramsey time increases, the density of the sample reduces due to two-body loss of molecules from the trap. We plot the detuning as a function of {\bf (a)} time and {\bf (b)} peak density. 
    Each interval is indicated by the $x$ error bars, with the marker indicating the centre of the interval. The dashed horizontal line is the detuning found when fitting across the whole dataset.  We see no evidence that the detuning is dependent upon the density.  
    }
    \label{fig:NoCollisionalShifts}
\end{figure*}

To look for evidence of collisional energy shifts we separate the results shown in Fig.~3 into six time-intervals, each 204\,ms long. We then fit the Ramsey fringes in each time-interval independently and plot the result in Supp. Fig.~\ref{fig:NoCollisionalShifts}(a). Over the 1200\,ms we interrogate the sample, the number of molecules remaining drops to $0.23N_i$. We estimate that two-body loss will increase the temperature of the sample from $0.7$\,$\mu$K to $\sim1.0$\,$\mu$K over this time. The density $n$ of the sample therefore reduces by the fraction
\begin{equation}
\frac{n}{n_\mathrm{i}} = \frac{N}{N_\mathrm{i}} \left(\frac{T}{T_\mathrm{i}}\right)^{-3/2} \approx 0.23 \times \left(\frac{1.0}{0.7}\right)^{-3/2} = 0.13,
\end{equation}
where $n_\mathrm{i}$ is the starting density of the sample. We see no significant change in the detuning of the microwaves as the density of the sample reduces, as shown in Supp.~Fig.~\ref{fig:NoCollisionalShifts}(b) where we plot the results as a function of the molecular density.

\begin{table*}
\begin{tabularx}{\textwidth}{c|c p{3.2cm}|c p{3.2cm}|c p{3.2cm}|c|c}
\hline
\hline
$B\,(G)$   & $\ket{0}$ &      &  $\ket{1}$ &     & $\ket{2}$ &  &  $E_{02}/h$\,(kHz)  &  $E_{12}/h$\,(kHz)  \\  
\hline
\hline
110.1  &  $(0,4)_1$  & $\equiv \hspace{1cm} -0.487\ket{0,0,3/2,5/2}$ $-0.874\ket{0,0,1/2,7/2}$   & $(0,3)_0$   &  $\equiv \hspace{1cm} 0.833\ket{0,0,3/2,3/2}$ $-0.522\ket{0,0,1/2,5/2}$ $+0.183\ket{0,0,-1/2,7/2}$ & $(1,3)_2$  & $\equiv \hspace{1cm} 0.054\ket{1,1,3/2,1/2}$ $-.059\ket{1,1,1/2,3/2}$ $-0.294\ket{1,1,-0.5,5/2}$ $-0.318\ket{1,1,-3/2,7/2}$ $+0.272\ket{1,0,3/2,3/2}$ $+0.199\ket{1,0,1/2,5/2}$ $-0.169\ket{1,0,-1/2,7/2}$ $+0.523\ket{1,-1,3/2,5/2}$ $+0.625\ket{1,-1,1/2,7/2}$ & 980,246.295& 980,327.278  \\
\hline
145.2  &  $(0,4)_1$  & $\equiv \hspace{1cm} 0.39\ket{0,0,3/2,5/2}$ $+0.920\ket{0,0,1/2,7/2}$   &  $(0,3)_0$  &  $\equiv \hspace{1cm} -0.893\ket{0,0,3/2,3/2}$ $+0.435\ket{0,0,1/2,5/2}$ $-0.113\ket{0,0,-1/2,7/2}$ & $(1,3)_2$ & $\equiv \hspace{1cm} -0.036\ket{1,1,3/2,1/2}$ $+0.0287\ket{1,1,1/2,3/2}$ $+0.205\ket{1,1,-1/2,5/2}$ $+0.290\ket{1,1,-1.5,7/2}$ $-0.195\ket{1,0,3/2,3/2}$ $-0.171\ket{1,0,1/2,5/2}$ $+0.129\ket{1,0,-1/2,7/2}$ $-0.467\ket{1,-1,3/2,5/2}$ $+0.755\ket{1,-1,1/2,7/2}$& 980,259.464& 980,336.593  \\
\hline
154.5  &  $(0,4)_1$  & $\equiv \hspace{1cm} 0.372\ket{0,0,3/2,5/2}$ $+0.928\ket{0,0,1/2,7/2}$   &  $(0,3)_0$  &  $\equiv \hspace{1cm} 0.905\ket{0,0,3/2,3/2}$ $+0.415\ket{0,0,1/2,5/2}$ $+0.100\ket{0,0,-1/2,7/2}$ & $(1,3)_2$  & $\equiv \hspace{1cm} -0.032 \ket{1,1,3/2,1/2}$ $+0.023\ket{1,1,1/2,3/2}$ $+0.186\ket{1,1,-1/2,5/2}$ $+0.281\ket{1,1,-3/2,7/2}$ $-0.179\ket{1,0,3/2,3/2}$ $-0.162\ket{1,0,1/2,5/2}$ $+0.120\ket{1,0,-1/2,7/2}$ $-0.450\ket{1,-1,3/2,5/2}$ $-0.781\ket{1,-1,1/2,7/2}$ & 980,262.071  & 980,339.056  \\
\hline
181.6  &  $(0,4)_1$  & $\equiv \hspace{1cm} 0.321\ket{0,0,3/2,5/2}$ $0.947\ket{0,0,1/2,7/2}$   & $(0,3)_0$   &  $\equiv \hspace{1cm} 0.928\ket{0,0,3/2,3/2}$ $-0.365\ket{0,0,1/2,5/2}$ $-0.074\ket{0,0,1/2,7/2}$ & $(1,3)_1$ & $\equiv \hspace{1cm} -0.080\ket{1,1,3/2,1/2}$ $+0.219\ket{1,1,1/2,3/2}$ $+0.162\ket{1,1,-1/2,5/2}$ $-0.110\ket{1,1,-3/2,7/2}$  $-0.687\ket{1,0,3/2,3/2}$ $-0.045\ket{1,0,1/2,5/2}$ $-0.018\ket{1,0,-1/,7/2}$ $-0.540\ket{1,-1,3/2,5/2}$ $+0.375\ket{1,-1,1/2,7/2}$ & 980,182.991& 980,260.992 \\
\hline
217.4  &  $(0,4)_1$  & $\equiv \hspace{1cm} -0.269\ket{0,0,3/2,5/2}$ $-0.963\ket{0,0,1/2,7/2}$   & $(0,3)_0$   &  $\equiv \hspace{1cm} 0.949 \ket{0,0,3/2,3/2}$ $-0.312\ket{0,0,1/2,5/2}$ $+0.051\ket{0,0,1/2,7/2}$ & $(1,4)_3$ & $\equiv \hspace{1cm} -0.323 \ket{1,1,3/2,3/2}$ $-0.585\ket{1,1,1/2,5/2}$ $+0.355\ket{1,1,-1/2,7/2}$ $-0.092\ket{1,0,3/2,5/2}$ $+0.645\ket{1,0,1/2,7/2}$ $-0.048\ket{1,-1,3/2,7/2}$ & 980,427.082& 980,508.760\\
\hline
154.5  &  $(0,4)_1$  & $\equiv \hspace{1cm} 0.372\ket{0,0,3/2,5/2}$ $+0.928\ket{0,0,1/2,7/2}$   &  $(0,3)_0$  & $\equiv \hspace{1cm}  0.904\ket{0,0,3/2,3/2}$ $-0.415\ket{0,0,1/2,5/2}$ $+0.100\ket{0,0,-1/2,7/2}$ & $(1,4)_3$ &$\equiv \hspace{1cm} 0.478\ket{1,1,3/2,3/2}$ $+ 0.517\ket{1,1,1/2,5/2}$ $-0.455\ket{1,1,-1/2,7/2}$ $+0.109\ket{1,0,3/2,5/2}$ $-0.529\ket{1,0,1/2,7/2}$ $-0.067\ket{1,-1,3/2,7/2}$  & 980,399.341 & 980,476.326 \\
\hline
\hline
\end{tabularx}
\caption{\label{table:StateCompositions} Compositions of states used in this work, written in the uncoupled basis $\ket{N, M_N, m_\mathrm{Rb}, m_\mathrm{Cs}}$ with coefficients rounded to one part in 10$^3$. At each magnetic field, $\ket{1}$ is chosen to couple well to both $\ket{0}$ and $\ket{2}$. For the measurement at 154.5\,G presented in Fig.~1, $\ket{1}=(1,3)_2$. This state has very weak coupling on the $\ket{1}\leftrightarrow\ket{2}$ transition. The measurements in Fig.~2 and Fig.~3, are also at 154.5\,G but instead use $\ket{1}=(1,4)_3$, which has significantly better coupling on this transition, but requires the use of the more complicated pulse sequence shown inset in Fig.~3. The final two columns give the calculated transition frequencies in free space. }
\end{table*}

\end{document}